\documentclass{article}

\RequirePackage{amsthm,amsmath,amsfonts,amssymb}
\RequirePackage[authoryear]{natbib}
\RequirePackage[colorlinks,citecolor=blue,urlcolor=blue]{hyperref}
\RequirePackage{graphicx}
\usepackage[margin = 1in]{geometry}
\usepackage{subcaption}

\usepackage{times}
\usepackage{ifthen}   
\usepackage{booktabs}
\usepackage{graphicx}
\usepackage{lmodern}
\usepackage{color}
\usepackage{multirow}
\usepackage{longtable} 	
\usepackage{datetime}
\usepackage{xspace}
\usepackage{paralist} 
\usepackage{enumitem} 
\usepackage{algorithmic}
\usepackage[ruled,vlined]{algorithm2e}

\usepackage{rotating}
\usepackage{comment}
\usepackage{cleveref} 

\usepackage{titlesec}

\usepackage{mdframed}  
\usepackage{colortbl}

\usepackage{soul}

\makeatletter
\def\namedlabel#1#2{\begingroup
    #2%
    \def\@currentlabel{#2}%
    \phantomsection\label{#1}\endgroup
}
\makeatother

\DeclareMathOperator*{\argmin}{arg\,min}

\usepackage{bm} 

\newcommand{\Kb}{\mathbf{K}}

\newcommand{\xb}{\mathbf{x}}
\newcommand{\Xb}{\mathbf{X}}

\newcommand{\Yb}{\mathbf{Y}}

\newcommand{\yobs}{\mathbf{y}_{\text{obs}}}

\usepackage[dvipsnames]{xcolor}

\usepackage{tikz}
\usetikzlibrary{positioning}
\usetikzlibrary{calc}
\usetikzlibrary{shapes.geometric}
\usetikzlibrary{topaths}
\usetikzlibrary{external}
\usetikzlibrary{fadings,decorations.pathreplacing} 

\usepackage{pgfplots}
\usetikzlibrary{shapes,decorations.pathmorphing,patterns}
\pgfplotsset{compat=newest}
\usepgfplotslibrary{fillbetween}

\usepackage[flushleft]{threeparttable}

\title{Staying on Track: Efficient Trajectory Discovery with Adaptive Batch Sampling}
\author{
Arindam Fadikar\thanks{Corresponding author. Email: \texttt{afadikar@anl.gov}. ORCID: 0000-0001-7396-0350} \\
\textit{Argonne National Laboratory}
\and
Abby Stevens\thanks{Email: \texttt{abby.e.stevens@gmail.com}. ORCID: 0000-0003-1976-1806} \\
\textit{Unaffiliated}
\and
Micka\"el Binois\thanks{Email: \texttt{mickael.binois@inria.fr}. ORCID: 0000-0002-7225-1680} \\
\textit{INRIA}
\and
Nicholson Collier\thanks{Email: \texttt{ncollier@anl.gov}. ORCID: 0000-0002-2376-4156} \\
\textit{Argonne National Laboratory}
\and
David O'Gara\thanks{Email: \texttt{dogara@anl.gov}. ORCID: 0000-0002-1957-400X}\\
\textit{Argonne National Laboratory}
\and
Jonathan Ozik\thanks{Email: \texttt{jozik@anl.gov}. ORCID: 0000-0002-3495-6735} \\
\textit{Argonne National Laboratory}
}
\date{}
\begin{document}
\maketitle

\begin{abstract}

Bayesian optimization (BO) is a powerful framework for estimating parameters of expensive simulation models, particularly in settings where the likelihood is intractable and evaluations are costly. In stochastic models every simulation is run with a specific parameter set and an implicit or explicit random seed, where each parameter set and random seed combination generates an individual realization, or trajectory, sampled from an underlying random process. Existing BO approaches typically rely on summary statistics over the realizations, such as means, medians, or quantiles, potentially limiting their effectiveness when trajectory-level information is desired. We propose a trajectory-oriented BO method that incorporates a Gaussian process surrogate using both input parameters and random seeds as inputs, enabling direct inference at the trajectory level. Using a common random number approach, we define a surrogate-based likelihood over trajectories and introduce an adaptive Thompson Sampling algorithm that refines a fixed-size input grid through likelihood-based filtering and Metropolis-Hastings-based densification. This approach concentrates computation on statistically promising regions of the input space while balancing exploration and exploitation. We apply the method to stochastic epidemic models, a simple compartmental and a more computationally demanding agent-based model, demonstrating improved sampling efficiency and faster identification of data-consistent trajectories relative to  parameter-only inference. 
\end{abstract}

\section{Introduction}
\label{sec:introduction}
Epidemiological simulation models are increasingly recognized as indispensable tools for supporting public health decisions during crises (e.g.,~\cite{Ozik2021, eubank2004modelling}). Their use dates back to at least the early twentieth century, when researchers began modeling how infectious diseases including influenza and, more recently, COVID-19, propagate through contact networks~\citep{kermack1927contribution}. The most common types of these models, such as compartmental models and agent-based models (ABMs), each have a set of parameters governing disease dynamics. Some models are deterministic, producing identical outputs for a given set of parameter inputs. Others are stochastic, generating outputs based on both model parameters and random seeds, producing specific stochastic realizations, or trajectories, of random processes within the models. Estimating model parameters typically involves fitting simulated trajectories to observed data~\citep{chatzilena2019contemporary, toni2009approximate, pooley2015posterior}. 
In this paper we investigate a novel Bayesian optimization (BO) approach that seeks to align simulation outputs with empirical data at the \textit{trajectory} level. This contrasts with more common approaches that rely on summary statistics
over stochastic realizations.

Classical calibration approaches rely on maximum likelihood estimation or Bayesian inference, often facilitated by surrogate models such as Gaussian processes (GPs) to emulate the expensive simulator and reduce computational cost. BO builds on this idea by embedding the surrogate in a sequential decision framework to prioritize promising regions of the parameter space. Other methods such as Ensemble Kalman Filtering (EnKF)~\citep{evensen2003ensemble, iglesias2013ensemble}, the Partially Observed Markov Process (POMP) framework~\citep{king2016statistical}, Approximate Bayesian Computation (ABC)~\citep{beaumont2002approximate}, and simulation-based inference (SBI)~\citep{cranmerFrontierSimulationbasedInference2020,dyerBlackboxBayesianInference2024a}
 are also used extensively in calibrating simulation models. However, calibrating stochastic simulation models remains a fundamentally ill-posed problem~\citep{baker2022analyzing}. Unlike deterministic models, where each input maps to a fixed output, stochastic simulators produce outputs that are random draws from an unknown distribution. The same input parameter $\xb$ can yield divergent outcomes across simulation replicates, making it ambiguous what it means for parameters to match observed data. Nonetheless, in common practice, calibration involves aligning expected values or other distributional summaries of stochastic simulation trajectories with data.

For the purpose of illustration, let us consider the use of ABMs for modeling infectious diseases. In an ABM, each individual is represented by an autonomous agent with predefined attributes and behaviors~\citep{eubank2004modelling}. Agents interact over network structures that capture heterogeneous mixing patterns, and a single ABM run produces one possible mixing and transmission trajectory~\citep{ferguson2020report, chang2021modelling}. Stochasticity arises from probabilistic transmission events but also from structural variations such as dynamic contact patterns or initial conditions. This results in different outcomes even under identical model parameters. When stochasticity reflects intrinsic structural variation, knowing only the parameter values that match observations ignores specific mixing patterns underlying transmission, which may be essential for planning targeted interventions. From a statistical perspective, such aggregation over randomness induces a loss of identifiability. Hence, we argue that a principled solution should investigate the benefits of estimating both the parameters $\xb$ and specific stochastic realizations, denoted by random seeds $r$, which jointly produce trajectories consistent with observed data. Recovering this tuple $(\xb, r)$ then enables trajectory-level inference. Essentially, rather than characterizing stochasticity as an inconvenient model behavior to be averaged out or minimized in expectation, trajectory-level inference characterizes it as a feature that can aid in explanation of observed dynamics in conjunction with model parameters. This more deliberate treatment may also be interpreted as constructing a set of candidate realizations of a stochastic process that explain observed data instead of being interpreted as the estimated expected outcome under infinite realizations. Individual trajectories also enable the use of trajectory continuations when modeling epidemics over time with updating surveillance data, such as with sequential data assimilation approaches~\citep{spannaus2022inferring, fadikar2024towards}.  

Trajectory-oriented optimization (TOO) is a surrogate-based framework for calibrating stochastic simulations by jointly estimating the tuple $(\xb, r)$ with the use of a common random number GP (CRNGP)~\citep{fadikar2023TOO}. The CRNGP enables the treatment of the simulation as a deterministic function over the augmented input space $(\xb, r)$, enabling inference at the trajectory level rather than through distributional summaries. Building on the TOO framework, this paper focuses on improving its computational efficiency to enable faster identification of simulated trajectories that are consistent with the observed data. We extend the Thompson Sampling (TS) based BO algorithm by adaptively generating the parameter-seed space for faster trajectory exploration. We illustrate this approach using a Susceptible-Exposed-Infectious-Recovered (SEIR) model of COVID-19 transmission in Chicago, highlighting the efficiency gains over existing trajectory-finding methods. We also discuss ways in which high-performance computing (HPC) workflows can extend the applicability of such complex algorithms to more computationally demanding simulations.

The remainder of the paper is organized as follows. \Cref{sec:background} introduces GPs as surrogate models for computationally expensive simulations and describes BO using TS as a sequential design strategy. This section also outlines the TOO framework. \Cref{sec:BO-TS} details the proposed adaptive-grid TS methodology, which integrates the CRNGP surrogate with an iterative refinement strategy to efficiently identify parameter–trajectory pairs consistent with observed data. In \Cref{sec:simulation}, we present a controlled simulation study comparing our trajectory-oriented methods with conventional parameter-focused approaches, highlighting the advantages of explicitly modeling stochastic realizations via CRNGP. Then, in \Cref{sec:usecase}, we apply the proposed methodology to a large-scale agent-based model of COVID-19 transmission, and demonstrate its practical utility in realistic scenarios for supporting public health stakeholders. We conclude in \Cref{sec:conclusion} with potential extensions of this work.


\section{Background}
\label{sec:background}

This section outlines the key components required for trajectory-level inference in stochastic simulation models. We begin by introducing GP surrogates, followed by strategies for incorporating input-dependent variability into the surrogate framework. We then describe a BO approach based on TS, which facilitates efficient exploration of the augmented $(\xb, r)$ input space.

\subsection{Gaussian Process Surrogates}

GPs~\citep{Rasmussen2005, gramacy2020surrogates} are nonparametric regression models used to infer unknown functions from observed data. GPs have been used extensively as surrogates~\citep{Sacks1989a, higdon2004combining} for expensive deterministic computer simulations, for tasks such as sensitivity analysis~\citep{saltelli2008global}, optimization~\citep{pearce2022bayesian}, calibration~\citep{Kennedy2001}, and uncertainty quantification~\citep{binois2015quantifying}. Formally, a GP defines a distribution over functions $ f: \mathbb{R}^d \to \mathbb{R} $, such that for any finite set of input locations $ \{\xb_1, \dots, \xb_N\} \subset \mathbb{R}^d $, the corresponding function evaluations $ \{f(\xb_1), \dots, f(\xb_N)\} $ follow a multivariate normal distribution. As a prior over functions, a GP is fully characterized by a mean function $ \mu(\cdot) $ and a positive semi-definite covariance kernel $ k(\cdot, \cdot) $, which encode assumptions about smoothness, stationarity, and correlation structure of the underlying function. Let $ \mathcal{D} = \{ (\xb_i, y_i) \}_{i=1}^N $ denote a training dataset consisting of $ N $ pairs of inputs $ \xb_i \in \mathbb{R}^d $ and corresponding scalar outputs $ y_i \in \mathbb{R} $, where we assume $ y_i = f(\xb_i) + \epsilon_i $ and $ \epsilon_i \sim \mathcal{N}(0, \tau^2) $ represents independent observation noise. Then, under a GP prior, the vector of outputs $ \Yb = (y_1, \dots, y_N)^\top $ is jointly Gaussian:
\begin{equation}
    \Yb \sim \text{MVN}(\boldsymbol{\mu}, \Kb_N + \tau^2 \mathbf{I}),
\end{equation}
where $ \boldsymbol{\mu} = (\mu(\xb_1), \dots, \mu(\xb_N))^\top $, and $ \Kb_N $ is the $ N \times N $ covariance matrix with entries $ k(\xb_i, \xb_j) $. The posterior predictive distribution at a new input $ \xb \in \mathbb{R}^d $ is also Gaussian, with conditional mean and variance given by:
\begin{align}\label{eq:kriging}
    \mu(\xb) &= \Kb(\xb)^\top (\Kb_N + \tau^2 \mathbf{I})^{-1} \Yb, \\
    \sigma^2(\xb) &= k(\xb, \xb) + \tau^2 - \Kb(\xb)^\top (\Kb_N + \tau^2 \mathbf{I})^{-1} \Kb(\xb),
\end{align}
where $ \Kb(\xb) = (k(\xb, \xb_1), \dots, k(\xb, \xb_N))^\top $ is the vector of covariances between the test point $ \xb $ and the training inputs. For notational convenience, we assume a zero mean function $ \mu(\cdot) = 0 $ in what follows. 

\subsection{Tracking the Input Dependent Variability}

While standard GP is a stationary process, and assumes homoskedastic noise, stochastic simulation models exhibit input-dependent variability that must be explicitly modeled. When replicate observations (multiple runs at the same design point) are available, the standard GP can be extended by aggregating replicate data and incorporating heteroskedastic noise into the covariance structure. One such extension is stochastic kriging~\citep{ankenmanstochastickriging}, which allows for input-dependent noise modeling by incorporating known or estimated variance at each input. This approach explicitly accounts for variability in simulation outputs at different input locations. An advancement over stochastic kriging is the heteroskedastic Gaussian Process (hetGP) model~\citep[e.g.,][]{binois2018heteroscedastic}, which jointly models both the mean and variance processes within a hierarchical framework. Unlike stochastic kriging, hetGP treats the variance as a latent function and estimates it from data, allowing for more flexibility in capturing input-dependent noise. However, it relies on the assumption that replicate variability can be adequately captured by a Gaussian distribution.

An alternative approach, the quantile Gaussian Process (QGP)~\citep{Plumlee2014, Fadikar2018}, models conditional quantiles of the output distribution rather than its mean or variance. QGP makes no distributional assumptions and provides a richer characterization of uncertainty, particularly when simulation outputs exhibit asymmetry or heavy tails. These methods differ in their assumptions and objectives: stochastic kriging incorporates known variance estimates, hetGP infers them as latent processes under Gaussian assumptions, and QGP directly targets the shape of the output distribution.

In contrast to these approaches, our work focuses on emulating individual simulation trajectories by treating the random seed $r$ as an additional input to the surrogate. This allows the surrogate to make predictions at the replicate level while simultaneously exploring the similarity among replicates across the input parameter space, with the goal of finding parameter-seed pairs $(\xb, r)$ that best reproduce observed data.

\subsection{Motivation for Trajectory-oriented Approaches}
Outcome variation for stochastic simulation models is driven by both changes to the model parameters as well as stochasticity. To explicate these two ideas, we first consider a simple example of an agent-based SIR model on a discrete two-dimensional grid with dimensions $[0,50]^2$. Agents move randomly on the grid, and Infected agents may infect Susceptible neighboring agents with probability $\beta$ and remain infected for $\gamma$ time steps. In some sense, this is a similar model to the well-known SIR system of ordinary differential equations (which is one form of a wider class of compartmental models). If we vary the per-contact transmission probability $\beta$ (which serves a similar role to the infection rate parameter in the compartmental approach), we expect a higher level of disease spread in the population.  However, the agent-based simulation has an important caveat: it does not assume a ``well-mixed'' population like the compartmental model, which leads to path-dependence in the model outcomes~\citep{hammond2015appendix}: an agent can only become Infected if one of its neighbors was already Infected at the previous time step, creating a chain of infection events. This also means that the $\beta$ parameter interacts nonlinearly with the agents' movements. Therefore, there are likely some specific infection parameter values that create just enough infections to lead to disease outbreaks for some random seeds, but die off in other cases. 

\begin{figure}[h!]
    \centering
    \includegraphics[width=\linewidth]{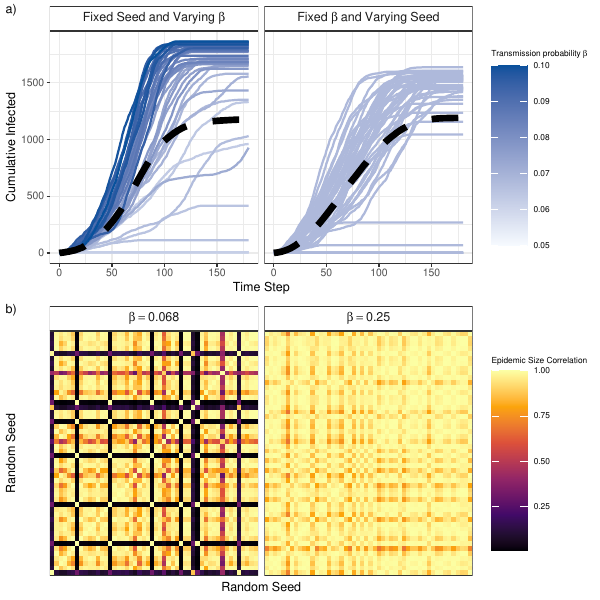}
    \caption{Variation in simulation outcomes due to changes in model parameters as well as stochasticity. Figure (a) shows cumulative epidemic curves for a simple SIR model of $2000$ agents on a grid, where the left-hand panel varies the $\beta$ parameter and fixes the random seed, while the right-hand panel fixes $\beta$ and varies the seed. Black dashed lines indicate the mean across trajectories. The red line in the left-hand panel denotes the trajectory for $\beta=0.068$, which is the fixed $\beta$ in right-hand panel. Figure (b) shows the squared correlation of cumulative infection curves across random seeds for two different values of $\beta$: when $\beta=0.068$ there is a high amount of variance between replicates, leading to widely different simulation outcomes, but when $\beta$ is much higher ($0.25$) the epidemic curves resemble each other much more closely across simulations. Each (x,y) point shows the squared correlation between two random seeds ordered from 1 to 50.}
    \label{fig:SIRGrid}
\end{figure}

To demonstrate this, we conduct the following experiment: we instantiate a population of $N=2000$ agents with $N-1$ non-infected (Susceptible) agents at unique locations on the grid and an index Infected agent near the top right corner.  We vary the per-contact transmission probability as well as the random seed, the results of which are shown in \Cref{fig:SIRGrid}. The top left-hand panel demonstrates the effect of increasing the $\beta$ parameter and how it leads to increased population-level infection, where the replicate is fixed across the $\beta$ values. We also note that if the model at hand was deterministic, this would be the only way to induce changes to the model outcomes. The top right-hand panel demonstrates the specific effect of stochasticity: for a fixed $\beta$ $=0.068$ (corresponding to the $\beta$ parameter of the red line in the left panel of \Cref{fig:SIRGrid}a), cumulative infection also varies widely \emph{across} random seeds. This highlights that in some simulations the epidemic fizzles out before infecting enough individuals to spread widely, and in some, takes off due to the idiosyncratic mixing that occurs in those specific replicates. These trajectories were selected so that each set induced a similar mean number of cumulative infections (shown as black dashed lines in both \Cref{fig:SIRGrid}a panels). Additionally, we see in the bottom panels (\Cref{fig:SIRGrid}b), where we track the correlations across random seeds, that the structure of the noise changes both across replicates and $\beta$ values. This supports the intuition that under certain conditions, information may be shared across stochastic replicates, in a similar fashion to how model parameters are learned through experimentation. This is relevant for active learning or BO workflows, where model runs are conducted sequentially, and whether a particular random seed was ``good'' or ``bad'' (the definition of which depends on the experiment's goal) would be valuable information for subsequent sampling of the parameter (and replicate) space.

\begin{figure}[!h]
    \centering
    \includegraphics[width = \textwidth]{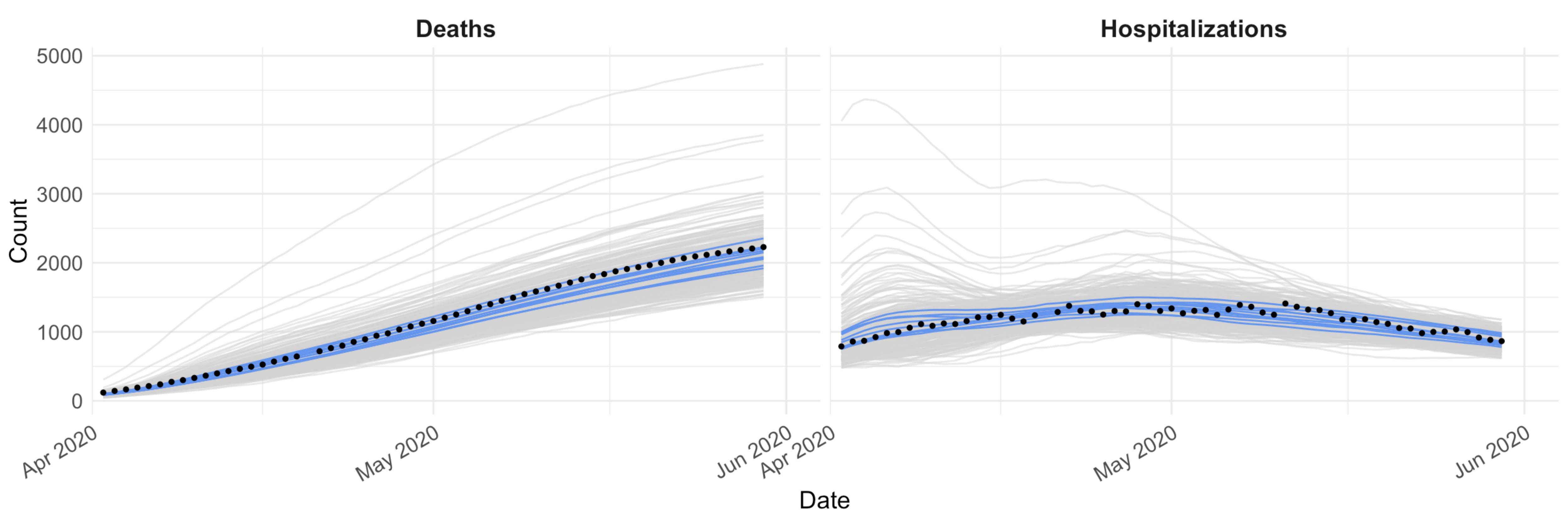}
   \caption{Cumulative deaths and incident hospitalizations from CityCOVID under 10 parameterizations identified in \cite{Ozik2021}. For each parameterization, 30 replicates (grey) and their mean (blue) are shown against observed data (black dots).}
    \label{fig:citycovid_top50}
\end{figure}

\Cref{fig:citycovid_top50} extends this intuition beyond a simple pedagogical model and applies it to a real-world model that informed high-stakes decision-making: it shows the calibrated deaths and hospitalizations trajectories from the CityCOVID model reported in \cite{Ozik2021}. For each input parameter, 30 stochastic replicates are shown (grey), together with their mean trajectory (blue). The figure illustrates that while mean trajectories can align well with the observed data (black dots), the individual realizations underlying these means can vary substantially, with many diverging away from the empirical outcomes. This variability highlights the influence of replicates in stochastic simulation models, where distinct realizations under identical parameters may produce very different epidemic trajectories. Consequently, calibration approaches that rely solely on matching mean behavior can conceal the presence of poor-quality trajectories and may fail to capture the specific realizations that are consistent with observed dynamics. This example serves as the primary motivation behind our goal of developing trajectory-oriented calibration approaches, which explicitly target the joint estimation of parameters and stochastic realizations, with the objective of identifying individual trajectories that align closely with the data. This also motivates us to look at the effectiveness of calibration algorithms and the quality of the discovered trajectories through the lens of \emph{errors} of the individual trajectories by the means of RMSE (root mean-squared error). More discussion follows in \Cref{sec:simulation_results}.  

\subsection{Common Random Number Gaussian Process (CRNGP)}

To enable trajectory-level inference over parameter-seed pairs, we adopt the Common Random Number Gaussian Process (CRNGP), a surrogate model designed for stochastic simulations~\citep{fadikar2023TOO}. The CRNGP models each simulation output using an augmented input pair $(\xb, r)$, where $\xb$ is the input parameter and $r$ is a replicate identifier (e.g., we use the random seed for initializing a simulation's random stream as the replicate identifier in this work), treated as a categorical variable~\citep{chen2012effects,pearce2019bayesian,pearce2022bayesian}. Formally, we define $f: \mathbb{R}^d \times \mathcal{S} \to \mathbb{R}$, where $\mathcal{S}$ denotes a countable set of replicates. The simulator is assumed deterministic over the joint input $(\xb, r)$, and a standard GP prior is placed on $f$ with a separable kernel structure:
\begin{equation}
  k((\xb_i, r), (\xb_j, r')) = k(\xb_i, \xb_j) \times k_s(r, r').
\end{equation}
The kernel $k$ encodes similarity between input locations and may follow standard choices (e.g., Gaussian, Mat\'ern). The replicate kernel $k_s$ models dependence between seeds. In the absence of known replicate structure, we assume:
\begin{equation}
  k_s(r, r') = \rho, \quad 0 < \rho < 1,
\end{equation}
which defines a constant overall similarity between different replicates. Simulation inputs and outputs are indexed as $\{(\xb_i, r_j)\}$ and $\{y_i^{(j)}\}$, where $r_i$ is the number of replicates at each $\xb_i$, and total simulations $N = \sum_{i=1}^n r_i$. Letting seeds take integer values, the full covariance matrix $\mathbf{K}_N$ has entries:
\begin{align}\label{eq:crncov}
    k((\xb_i, r), (\xb_j, r')) = \begin{cases}
        k(\xb_i, \xb_j), & \text{if } r = r', \\
        \rho \times k(\xb_i, \xb_j), & \text{if } r \ne r'.
    \end{cases}
\end{align}
The resulting kernel $k$ is positive definite and the inter-replicate similarity is controlled via $\rho$. A nugget term $\tau^2$ is added to the diagonal to form the full covariance $\mathbf{K}_N + \tau^2 \mathbf{I}$. The kriging equations from~\Cref{eq:kriging} remain applicable, and predictions can be made at any $(\xb, r)$. Averaging over seeds corresponds to predicting at a new, unobserved replicate~\citep{pearce2022bayesian}.

\subsection{Bayesian Optimization with Thompson Sampling}\label{sec:bo-ts}
Bayesian optimization is a sequential design framework for optimizing expensive black-box functions~\citep{garnett2023bayesian, Jones1998}. It proceeds by constructing a probabilistic surrogate model, such as a GP over the objective function, and uses posterior uncertainty to guide the selection of future evaluation points. 

Among acquisition strategies, Thompson Sampling (TS)~\citep{thompson1933likelihood} is a simple yet effective approach that draws a sample from the posterior distribution of the surrogate model and selects the input corresponding to the optimum of the sampled function. This randomized strategy implicitly balances exploration and exploitation. In each iteration, a surrogate $ f \sim \mathcal{GP}(\mu, k) $ is sampled over a candidate grid, and the next input is chosen as the optimizer of the sampled realization. 

In the context of stochastic simulations with multiple replicates at each input, the standard GP surrogate is insufficient for realization-level modeling, necessitating a surrogate like the CRNGP model introduced in the previous section. When combined with TS, this enables BO to operate directly at the level of simulated trajectories, allowing for discovery of ``good'' trajectories. We summarize the standard TS procedure using a CRNGP surrogate in Algorithm~\ref{algo:TS}, which serves as the foundation for the adaptive-grid TS method introduced in the next section.

\begin{algorithm}[!ht]
\caption{Thompson Sampling with CRNGP}\label{algo:TS}
\begin{algorithmic}[1]
\REQUIRE $N_0$ (initial budget), $N_{\max}$ (total budget), observed data $\yobs$, per TS iteration sample batch size $J$, the (parameter, replicate) candidate set $\tilde{\Xb} \times \mathcal{R}$
\STATE Generate initial design $\{(\xb_i, r_i)\}_{i=1}^{N_0}$ and compute simulation outputs $\{f(\xb_i, r_i)\}$
\STATE Define discrepancy function: $d_i = d(f(\xb_i, r_i), \yobs)$
\STATE Train CRNGP surrogate on inputs $\{(\xb_i, r_i)\}$ and responses $\{d_i\}$
\WHILE{$n < N_{\max}$}
    \STATE Sample $ \tilde{d}^{(j)} \sim \text{CRNGP posterior}, \;\; j = 1, \cdots, J $
    \STATE Select next batch of inputs $(\xb_{n+1}^{(j)}, r_{n+1}^{(j)}) \in \argmin_{(\xb, r) \in \tilde{\Xb} \times \mathcal{R}} \tilde{d}^{(j)}(\xb, r), \forall j$
    \STATE Evaluate simulator: $f(\xb_{n+1}^{(j)}, r_{n+1}^{(j)}), \forall j$
    \STATE Compute discrepancy for the new evaluations: $d^{(j)}_{n+1} = d(f(\xb^{(j)}_{n+1}, r^{(j)}_{n+1}), \yobs), \forall j$
    \STATE Augment training data and update CRNGP surrogate
    \STATE $n \leftarrow n + J$
\ENDWHILE
\end{algorithmic}
\end{algorithm}

\paragraph*{Motivation for Using Thompson Sampling.} 
The integration of TS with the CRNGP surrogate provides a principled strategy for addressing trajectory-level optimization in stochastic simulations. In our setting, the objective is not only to identify input parameters \(\mathbf{x}^*\) that minimize discrepancy with observed data \(\yobs\), but also to recover the specific stochastic realization \(r^*\) that yields a plausible match. This leads to a joint optimization problem of the form \((\mathbf{x}^*, r^*) \in \arg\min_{\mathbf{x}, r} d(f(\mathbf{x}, r), \yobs)\), where \(d(\cdot, \cdot)\) is a suitably defined discrepancy metric. Traditional GP surrogates that marginalize over simulation noise are ill-suited for this task, as they omit information about individual trajectories. The CRNGP surrogate, by explicitly modeling simulation outputs over the augmented input space \((\mathbf{x}, r)\), enables realization-level inference. TS then provides an efficient mechanism for navigating this augmented space by sequentially selecting candidate pairs \((\mathbf{x}, r)\) that minimize predicted discrepancy. This framework also naturally supports batch evaluation and parallel execution, making it well-suited for high-performance computing environments and time-sensitive applications such as real-time epidemiological forecasting.

\section{Thompson Sampling Over An Adaptive Grid}
\label{sec:BO-TS}
To support scalable BO for efficient trajectory finding, we adopt a discrete search strategy wherein TS is performed over a finite set of candidate input points. This grid-based formulation, as outlined in~\Cref{algo:TS}, offers a computationally tractable alternative to continuous optimization, especially in the context of simulation-based models where each function evaluation can be costly. By limiting evaluations to a fixed set of locations, the approach avoids expensive numerical optimization routines~\cite{agrawal2013thompson} and enables efficient batch evaluations in HPC settings~\cite{bergstra2012random, gonzalez2016batch}.

Grid-based TS has been widely used in multi-armed bandit problems~\cite{chapelle2011empirical}, hyperparameter tuning~\cite{snoek2012practical}, and the calibration of stochastic simulators~\cite{frazier2018tutorial, fadikar2023TOO}. While the fixed-grid approach simplifies implementation and maintains statistical efficiency, its performance depends heavily on the resolution and coverage of the search space. Coarse grids may miss important regions of the parameter space, particularly in high-dimensional or nonlinear settings. Recent developments address this limitation through adaptive discretization schemes. For instance,~\citet{binois2020kalai} propose iteratively refining the grid to enhance coverage, while~\citet{gramacy2022triangulation} use geometric structures to select informative candidate points. These strategies retain the computational advantages of grid-based search while enabling greater flexibility and improved optimization performance. In the next section, we describe our adaptive grid strategy tailored for trajectory-level optimization using CRNGP surrogates.

\subsection{Adaptive Grid Refinement Strategy for Efficient Thompson Sampling}

To address the limitations of searching over a fixed-grid, we introduce an adaptive grid refinement strategy that enhances TS performance by dynamically updating the search space at each iteration. This approach aims to focus computational effort on regions with high potential for matching observed data while avoiding unnecessary evaluations in less informative areas. This idea is conceptually related to strategies such as trust-region modeling and space partitioning, both of which adaptively localize the search. Trust-Region Bayesian Optimization (TuRBO)~\cite{eriksson2019scalable} confines the sampling to high-performing neighborhoods, whereas space-partitioning methods~\cite{bubeck2011bandit, wang2020lamcts} divide the parameter space and allocate more resources to promising subregions. Building on these ideas, we implement an intuitive and scalable alternative. Specifically, we design a two-stage refinement mechanism applied to the candidate grid $\tilde{\Xb}_t$ at each BO iteration $t$:  
\begin{enumerate}
    \item \textbf{Filtering}: Discard candidate points with low posterior likelihood of producing trajectories close to observed data based on previous evaluations, i.e., with high value of the discrepancy function $d$.
    \item \textbf{Densification}: Sample new points around the top-performing candidates using localized perturbations, guided by the GP prediction.
\end{enumerate}
The size of the grid remains fixed throughout the BO, while sequentially improving the resolution near high-quality regions. In the context of trajectory-based calibration for epidemiological simulations, we show below that this enables more efficient identification of both optimal parameter configurations and corresponding stochastic realizations, facilitating fast time-to-solution.

\subsubsection{Filtering Step}\label{sec:filter}

We begin each TS iteration by generating a set of candidate points $\tilde{\Xb}^{(0)}_t = \{(\xb_j, r_j)\}_{j=1}^M$\, using a uniform Latin Hypercube Sampling (LHS) design over the joint input space of model parameters $\xb$ and trajectory identifiers $r$. To evaluate which candidate points are most likely to produce simulated outputs that match the observed data $\yobs$, we define a likelihood function based on the discrepancy between simulated and observed outcomes. A CRNGP surrogate is trained to emulate $d(f(\xb, r), \yobs)$ using the already evaluated simulations $\mathcal{D}_{t-1}$ until the previous TS iteration. Let $\tilde{d}(\xb, r \mid \mathcal{D}_{t-1})$ denote a sample from the surrogate’s predictive distribution at input $(\xb, r)$. We then define the likelihood function as:
\begin{equation}\label{eq:likelihood}
L\left((\xb, r) \mid \yobs, \mathcal{D}_t\right) = p_d\left(\tilde{d}(\xb, r \mid \mathcal{D}_t)\right),
\end{equation}
where $p_d$ is a probability density function centered at zero. A natural choice is $p_d(z) = \mathcal{N}(z; 0, \sigma_{\text{obs}}^2)$, allowing for noise in the observed data.

We perform importance resampling on the LHS grid $\tilde{\Xb}_t$, treating it as a sample from a uniform prior over $(\xb, r)$. Each candidate is assigned an importance weight equal to its likelihood value:
\[
w(\xb, r) = L\left((\xb, r) \mid \yobs, \mathcal{D}_t\right).
\]
A new refined grid $\tilde{\Xb}^{F}_{t}$ is obtained by resampling $(\xb, r)$ pairs from the initial grid according to their normalized importance weights, which discards points with low predictive likelihood and preserves high-likelihood regions for further refinement.

\subsubsection{Densification Step}\label{sec:dense}

After filtering, the updated grid $\tilde{\Xb}^F_{t}$ may contain fewer than $M$ total points. To maintain a consistent grid size across TS iterations, we densify the search set using a Metropolis-Hastings (MH)-inspired sampling strategy. Unlike standard MH algorithms, here we operate in the augmented input space $(\xb, r)$ with the restriction that proposals are generated solely in the continuous parameter space $\xb \in \mathcal{X}$, while the set of random seeds $\mathcal{R} = \{r_1, \dots, r_l\}$ remains fixed, with total number of replicates $l$, throughout the BO procedure. For each proposed parameter value $\xb_{\text{can}} \sim q(\cdot \mid \xb)$, we construct the set of candidate inputs
\[
\{(\xb_{\text{can}}, r_1), \dots, (\xb_{\text{can}}, r_l)\}
\]
by pairing the proposed $\xb_{\text{can}}$ with each seed value in $\mathcal{R}$. Each pair is evaluated using the surrogate likelihood $L\big((\xb_{\text{can}}, r_j) \mid \yobs, \mathcal{D}_t, \cdot\big)$ defined via the CRNGP on the difference function. One such pair is accepted into the refined grid using the MH acceptance probability:
\begin{equation}
  \alpha = \min\left\{1, \frac{L\big((\xb_{\text{can}}, r_j) \mid \yobs, \mathcal{D}_t, \cdot\big) \cdot q(\xb_{\text{can}} \mid \xb)}{L\big((\xb, r) \mid \yobs, \mathcal{D}_t, \cdot\big) \cdot q(\xb \mid \xb_{\text{can}})}\right\}.  
\end{equation}
This process is repeated until the total number of unique points in $\tilde{\Xb}_{t+1}$ reaches $M$. \Cref{algo:grid} provides the pseudo code for the adaptive grid generation process.

\begin{figure}[!htbp]
    \centering
    \includegraphics[width=1\linewidth]{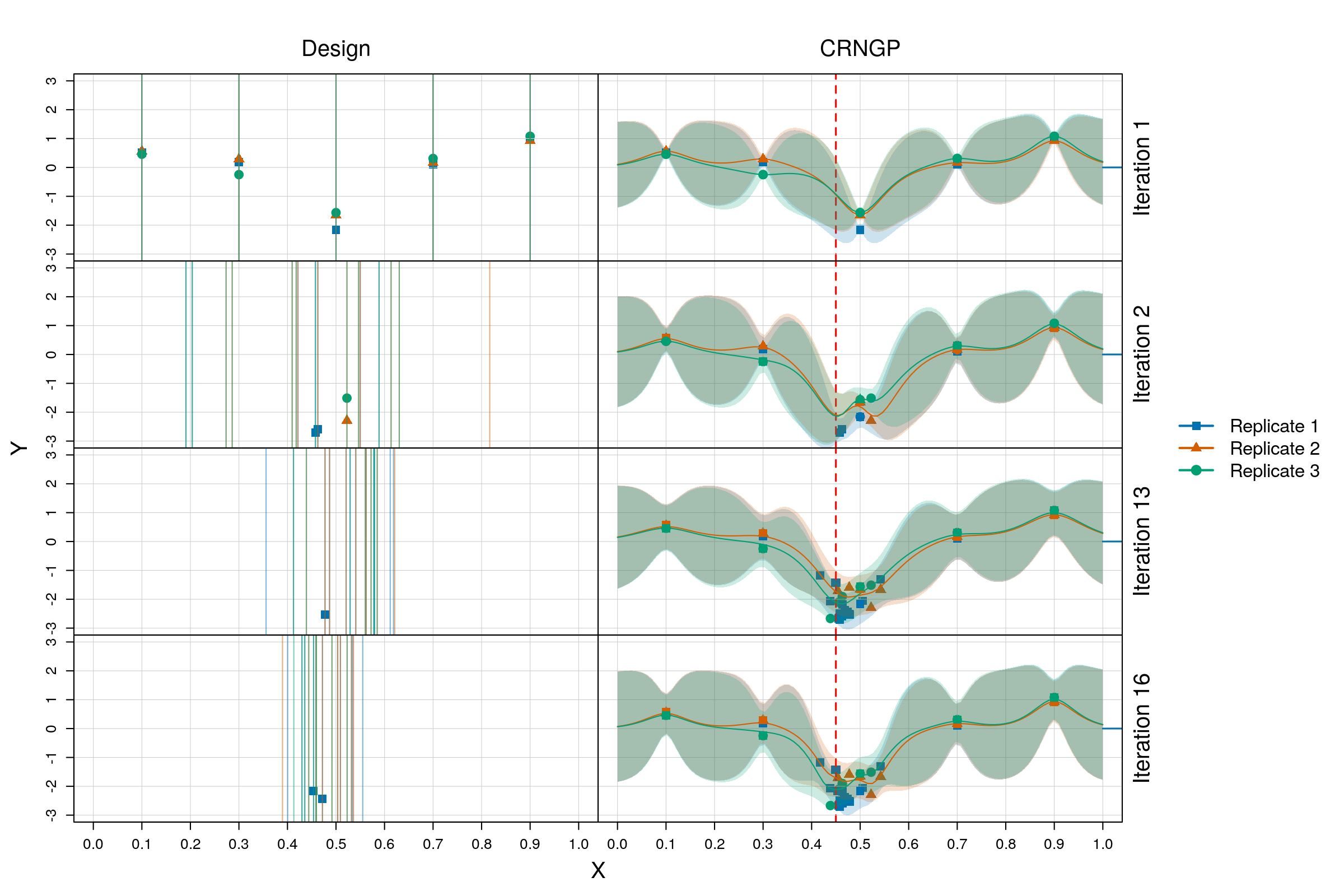}
   \caption{Illustration of adaptive grid TS with a CRNGP surrogate to find $(X, r)$ tuple having minimal $Y$ values. The example consists of three replicates denoted by different colors. Left panels show the evolving grid locations in the augmented input space, with colored vertical lines indicating candidate input locations (and replicate) at each BO iteration. Among those candidate locations, the evaluated points are indicated by corresponding plotting characters. Right panels show the CRNGP posterior mean (solid lines) and uncertainty bands for each replicate using the evaluated points until the corresponding iteration. The red dashed vertical line indicates the parameter value at which the ground-truth is generated. From top to bottom, rows correspond to increasing BO iterations. }
    \label{citeexample-BO}
\end{figure}

\Cref{citeexample-BO} illustrates how adaptive grid TS combined with a CRNGP surrogate refines the search for parameter-seed tuples $(X, r)$ that minimize the discrepancy $Y = d(f(X, r), \yobs)$. The example is based on the stochastic SIR model described in \cref{sec:simulation}, and is constructed in a one-dimensional setting by fixing the recovery rate at $\gamma = 0.35$ (normalized to the $[0,1]$ scale) and varying the transmission rate $X = \beta$. Three stochastic replicates are considered, corresponding to distinct random seeds. Synthetic ground truth observation is generated by setting $\beta = 0.45,\; r = 5$, and the objective of the optimization is to identify pairs $(X, r)$ for which the simulated trajectory closely matches the observed data, i.e., to have minimum $Y$ values. The discrepancy values are log-transformed and standardized to have mean zero and unit variance prior to fitting the surrogate and applying BO.

The left panels show the evolution of the candidate grid across successive BO iterations. Vertical colored lines represent the set of candidate input locations considered at each iteration, and the plotting symbols indicate the subset of locations that are evaluated. As the iterations progress, the grid increasingly concentrates around regions associated with lower predicted discrepancy values as a result of likelihood-based filtering and grid densification. The right panels display the corresponding CRNGP posterior summaries for each replicate, where solid lines denote posterior means and shaded regions indicate posterior uncertainty. As additional evaluations are incorporated, uncertainty decreases in the region of interest (near $X = 0.45)$. This simple 1-D example demonstrates how the CRNGP surrogate enables replicate-aware inference, and how adaptive grid refinement guides exploration toward regions of the augmented input space that are most likely to yield high-quality trajectories.

\begin{algorithm}[!ht]
\caption{Pseudo-code for Adaptive-grid generation with fixed seed set $\mathcal{R}$}\label{algo:grid}
\begin{algorithmic}[1]
\REQUIRE  $\tilde{\Xb}^{(0)}_t = \{(\xb_1, r_1), \dots, (\xb_M, r_M)\}$ (grid of size $M$), fixed seed set $\mathcal{R} = \{r_1, \dots, r_l\}$

\FOR{$i = 1, \dots, M$}
    \STATE $w(\xb_i, r_i) \gets L\big((\xb_i, r_i) \mid \yobs, \mathcal{D}_t, \cdot\big)$
    \STATE $w'(\xb_i, r_i) \gets w(\xb_i, r_i) / \sum_{j=1}^M w(\xb_j, r_j)$
\ENDFOR
\STATE Resample $(\xb_i, r_i)$ with probability $w'(\xb_i, r_i)$ (with replacement) to obtain $\tilde{\Xb}^F_{t}$
\STATE Initialize $\tilde{\Xb}_{t}^{\text{adapt}} \gets \tilde{\Xb}^F_{t}$
\STATE $m \gets |\tilde{\Xb}_{t}^{\text{adapt}}|$
\WHILE{$m < M$}
    \STATE Choose $(\xb, r) \in \tilde{\Xb}_{t}^{\text{adapt}}$ at random
    \STATE Sample $\xb_{\text{can}} \sim q(\cdot \mid \xb)$
    \FOR{$r_j \in \mathcal{R}$}
        \STATE Compute $\alpha = \min\left\{1, \frac{L\big((\xb_{\text{can}}, r_j) \mid \yobs, \mathcal{D}_t, \cdot\big) \cdot q(\xb_{\text{can}} \mid \xb)}{L\big((\xb, r) \mid \yobs, \mathcal{D}_t, \cdot\big) \cdot q(\xb \mid \xb_{\text{can}})}\right\}$
        \STATE Sample $u \sim \text{Uniform}(0, 1)$
        \IF{$u < \alpha$ and $(\xb_{\text{can}}, r_j) \notin \tilde{\Xb}_{t}^{\text{adapt}}$}
            \STATE $\tilde{\Xb}_{t}^{\text{adapt}} \gets \tilde{\Xb}_{t}^{\text{adapt}} \cup (\xb_{\text{can}}, r_j)$
            \STATE $m \gets m + 1$
            \STATE \textbf{break}
        \ENDIF
    \ENDFOR
\ENDWHILE
\STATE Return $\tilde{\Xb}_{t}^{\text{adapt}}$ as the updated search grid
\end{algorithmic}
\end{algorithm}

\paragraph*{Fast Sampling from Conditional GP}
Step 5 of~\Cref{algo:TS} requires sampling from the GP posterior on the $M$ candidates. For large $M$, it may become computationally demanding, with a $\mathcal{O}(M^3)$ complexity. Assuming that all seeds are considered across the unique $\xb$'s among the $M$ candidates, the resulting virtual candidate set with $\tilde{M}$ candidates leads to a covariance matrix over all possible ($\xb,r$) combinations. Given the separable covariance structure between the $\xb$'s and $r$'s, this covariance matrix has a Kronecker structure, leading to a faster Cholesky decomposition of the covariance over all $\xb$'s and over all $r$'s separately.
Then, based on the technique described by \cite{chiles2012geostatistics,chevalier2015fast}, these unconditional simulations can be conditioned by subtracting the GP prediction on this realization before adding the prediction on the actual data.


\section{Application of Adaptive Grid TS and Performance Comparisons}
\label{sec:simulation}

In this section, we demonstrate the advantages of using both a CRNGP surrogate and an adaptive grid sampling procedure (\textbf{aCRN}) within the TS framework to efficiently identify high-quality trajectories. We compare the performance of \textbf{aCRN} to other surrogate-based TS sampling methods through experiments that leverage simulations from a stochastic epidemiological Susceptible-Infectious-Recovered (SIR) compartmental model. The model, which is described in detail below, has two parameters, infection rate ($\beta$) and recovery rate ($\gamma$). For a given pair of parameters, $(\beta, \gamma)$, we generate a single disease \textit{trajectory} by specifying the random seed $r$. We select a parameter-seed combination $(\beta^*, \gamma^*, r^*)$ to simulate a ground-truth trajectory $\mathbf{y}_{\text{obs}}$, and compare the quality of trajectories each TS method finds for a given simulation budget.

\subsection{Model}

We generate synthetic outbreaks with a discrete–time stochastic
SIR model implemented using the \texttt{odin} R package~\cite{fitzjohn2024odin}. In a typical SIR model, a population is partitioned into \textit{compartments}, and the following system of ordinary differential equations describes how individuals transition between compartments:
\begin{equation}\label{eq:sir}
    \frac{dS}{dt} = -\beta \frac{SI}{N}, \quad\quad \frac{dI}{dt} = \beta \frac{SI}{N} - \gamma I, \quad\quad \frac{dR}{dt} = \gamma I,
\end{equation}
where $S$, $I$, and $R$ represent the number of individuals who are susceptible, infected, and recovered, respectively; $N = S + I + R$ is the total population size, $\beta$ is the infection rate, and $\gamma$ the recovery rate. For small time steps, we can discretize this system as:
\begin{equation}
    S_{t+1} =  S_t -\beta \frac{S_tI_t}{N_t}, \quad\quad I_{t+1} = \beta \frac{S_tI_t}{N_t} - \gamma I_t, \quad\quad R_t = \gamma I_t.
\end{equation}
The deterministic transition updates standard in SIR models do not capture the variability and uncertainties inherent to real-world infection dynamics, so we instead model the transition rates probabilistically. Specifically, we implement and use the following transition equations in our model, which allow for randomness in the transition between compartments:
\begin{equation}\label{eq:sir_stoch}
    S_{t+1} = S_t - X_t, \quad\quad
I_{t+1} = I_t + X_t - Y_t, \quad\quad
R_{t+1} = R_t + Y_t, 
\end{equation} 
where,
$$X_t \sim \mathrm{Binom}\left(S_t,\;1-\exp\left(-\beta \frac{I_t}{N}\right)\right), \quad \quad Y_t \sim \mathrm{Binom}\left(I_t,\; 1-\exp\left(-\gamma\right) \right)$$

For all experiments, we run the simulation for $T=100$ time steps, and we set $N = 1010$ as the total population size, with most individuals in the susceptible compartment ($S_0= 1000$), several individuals in the infected compartment ($I_0 = 10$), and no recovered individuals ($R_0=0$). In addition to accepting values for $\beta$ and $\gamma$, our implementation accepts a random seed $r$ that fixes the random stream for each model run.

\subsection{Experiments}
Our experiments explore the conditions under which we can find trajectories sufficiently close to the ground-truth within a fixed simulation budget $N_{max}$. We initialize each optimization algorithm we evaluate with a space-filling design of $n_{init}$ model parameters, each of which is replicated $n_{rep}$ times for a total initial cost of $N_0 = n_{init}\cdot n_{rep}$ simulation runs. Let $\mathbf{y}_{ij} = f_T(\beta_i, \gamma_i, r_j)$ represent the $T$-dimensional trajectory output of the SIR model, evaluated at $(\beta_i, \gamma_i)$ with the random seed set to $r_j$. Given the ground-truth trajectory $\mathbf{y}_{\text{obs}}$, $z_{ij} := \| \mathbf{y}_{ij} - \mathbf{y}_{\text{obs}}\|_2^2$ is the scalar output for training the GP surrogates on and the quantity we optimize over. After the GP surrogate is fit on the initial design, additional $(\beta, \gamma, r)$ locations are selected to evaluate through batched TS algorithms until a total of $N_{max}$ simulations are run. Each iteration of the TS procedure selects $n_{TS}$ new points to evaluate from a grid (either fixed or adaptive) of size $M$. Each of these hyperparameters ($N_{max}$, $n_{init}$, $n_{rep}$, $n_{TS}$, $M$) has the potential to impact the performance of the algorithms, and we perform experiments for all combinations (see \Cref{table:exp_params}), replicating each experiment 10 times to account for algorithmic stochasticity (note this is separate from specifying the simulation random seed) for a total of 990 experimental runs. 

\begin{table}[h!]
\centering
\begin{tabular}{l l l} 
 \hline
 \textbf{Hyper-parameter} & \textbf{Description} & \textbf{Values} \\ [0.5ex] 
 \hline
$N_{max}$ & Total simulation budget & 300, 500, 700 \\
$n_{init}$ & Number of unique $(\beta, \gamma)$ in initial design &  5, 10 \\
$n_{rep}$ & Number of replicates in initial design & 10, 20 \\
$n_{TS}$ & Number of trajectories sampled per TS iteration & 10, 20, 30 \\
$M$ & Number of points in grid & 100, 200, 300 \\ 
 \hline
\end{tabular}
\vspace{0.2cm}
\caption{Optimization algorithm hyper-parameters experimental design}
\label{table:exp_params}
\end{table}

In each experimental setting, we seek to identify trajectories that are close to the ground truth using our method (\textbf{aCRN}) and four comparators, each of which combines a different surrogate model and grid sampling scheme (\Cref{table:comparators}). The key difference between CRNGP and hetGP surrogates is their treatment of replicates. CRNGP considers replicates as model inputs and hetGP seeks to understand the mean behavior from additional replicates at a given $(\beta, \gamma)$. In our implementation of \textbf{aHet} and \textbf{fHet}, parameter locations are selected based on $(\beta, \gamma)$ alone, and once a new parameter is selected, a random seed is randomly assigned to generate a new trajectory. In the CRNGP-based methods, however, the random seed becomes part of the parameter space. For \textbf{aCRN} and \textbf{fCRN}, the universe of random seeds is fixed to the $n_{rep}$ seeds included in the initial design. Early experimentation revealed that when fixing both the grid and random seeds, as we do in \textbf{fCRN}, we quickly exhaust available sampling locations, as we cannot continue assigning additional random seeds to a given $(\beta, \gamma)$ as we can with hetGP. Because of this, we include an additional method, \textbf{fgCRN}, that allows flexible assignment of random seeds to the fixed grid - that is, if a particular $(\beta, \gamma)$ location has already been evaluated $n_{rep}$ times, we allow additional random seeds outside the initial set to be evaluated for that point.

\begin{table}[h!]
\centering
\begin{tabular}{l l l l} 
 \toprule
 \textbf{Method} & \textbf{Surrogate} & \textbf{Grid Sampling} & \textbf{Random Seeds} \\ [0.5ex] 
 \midrule
 \textbf{aCRN} & CRNGP & Adaptive & Fixed \\ 
  \textbf{fCRN}  & CRNGP & Fixed & Fixed \\
  \textbf{fgCRN}  & CRNGP & Fixed & Adaptive \\
  \textbf{aHet}  & hetGP & Adaptive & Flexible \\
  \textbf{fHet}  & hetGP & Fixed & Flexible \\ [1ex] 
 \bottomrule
\end{tabular}
\caption{Comparator approaches for TS-based BO}
\label{table:comparators}
\end{table}

These experiments were performed on Argonne Laboratory Computing Resource Center's Improv cluster using the EMEWS framework~\cite{ozikDesktopLargeScaleModel2016a,collierDistributedModelExploration2024}. All BO algorithms were implemented in R. All the code used in this paper is publicly available at \url{https://github.com/emews/adaptiveTS}.

\subsection{Results}
\label{sec:simulation_results}



\begin{figure}[!h]
    \centering
    \includegraphics[width=0.9\textwidth]{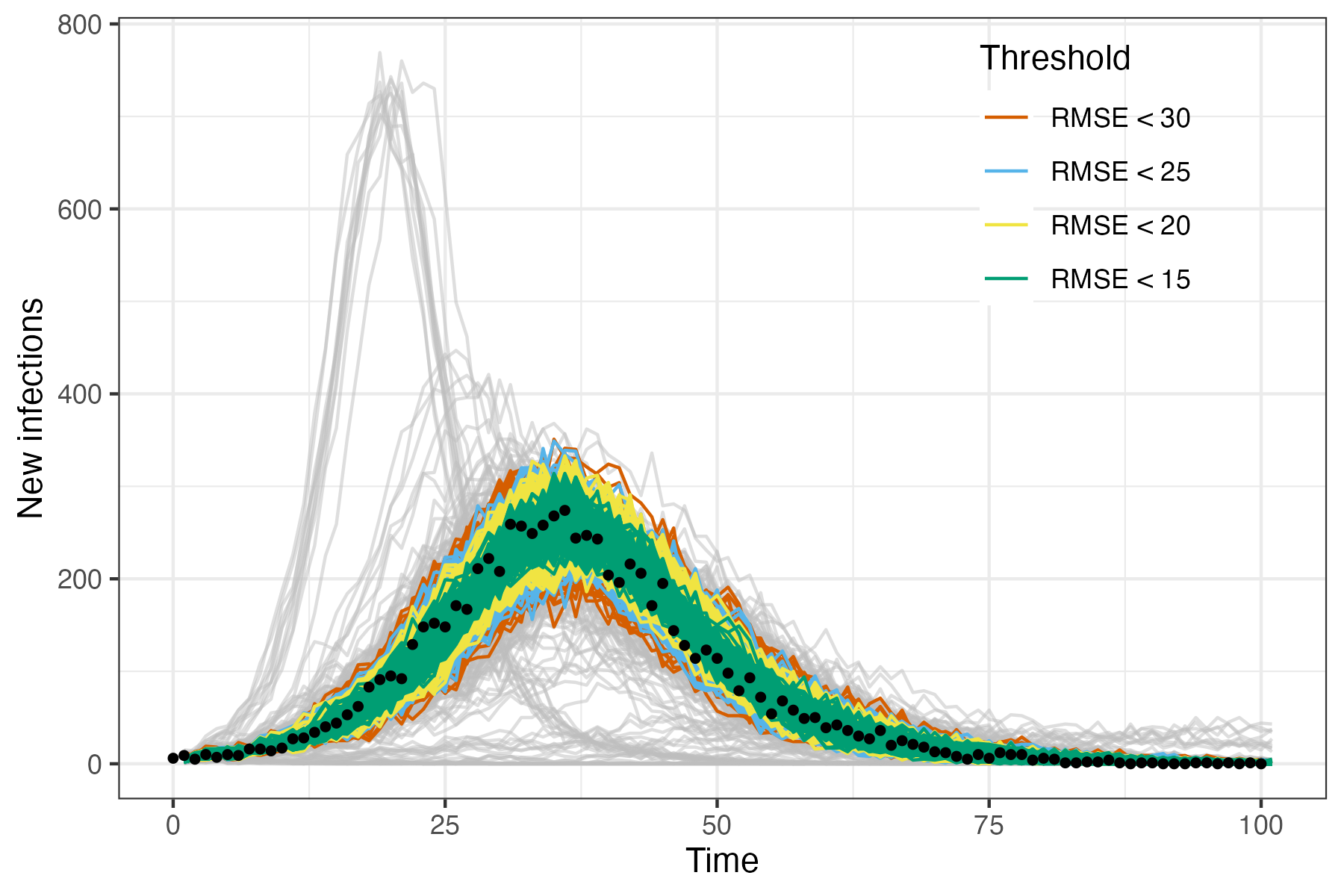}
    \caption{Simulated epidemic trajectories found by the aCRN method for a single experiment configuration. Each trajectory is colored according to its RMSE category, with thresholds at 15, 20, 25, and 30. The ground truth is shown using black dots.}
    \label{fig:thresholds}
\end{figure}

To evaluate both the speed and overall utility of each method in finding high-quality model trajectories, we approach our analysis through a \textit{thresholding} lens. \Cref{fig:thresholds} shows all the trajectories explored by \textbf{aCRN} for a particular experiment configuration, colored by a set of RMSE threshold categories with respect to a simulated ground truth (black dots). We note that acceptable thresholds will vary based on stakeholder needs and use-case. 

\subsubsection{Collecting High-Quality Trajectories}
\begin{figure}[!h]
    \centering
    \includegraphics[width=\textwidth]{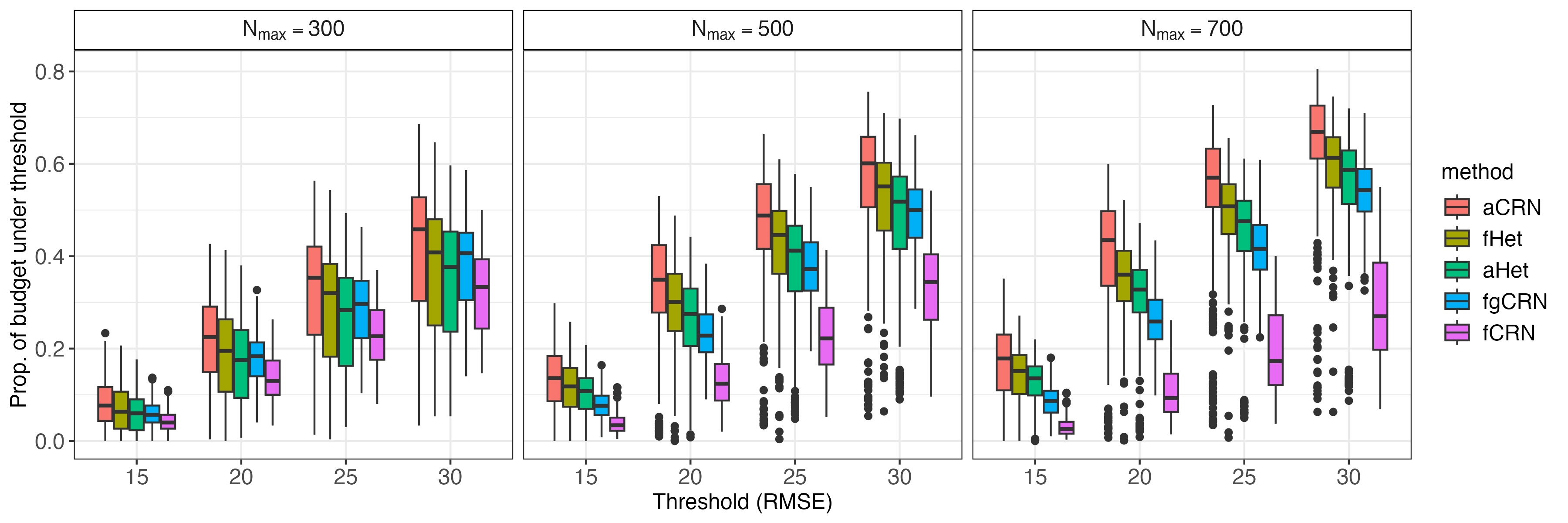}
    \caption{Proportion of trajectories with RMSE below varying thresholds across all experiments varying the hyper-parameters in~\Cref{table:exp_params}, stratified by simulation budget $N_{\max}$.}
    \label{fig:overall}
\end{figure}

We first assess the overall quality of trajectories explored by each method in a given experiment. For a given threshold $t$, we compute the proportion of the simulation budget ($N_{max}$) trajectories with RMSE less than $t$, with a high proportion indicating an efficient use of the simulation budget. \Cref{fig:overall} shows the distribution of this metric across all experiments, stratified by $N_{max}$. \textbf{aCRN} consistently identifies the highest proportion of trajectories under RMSE thresholds relative to comparators, across all thresholds and simulation budgets. We observe the relative benefits of \textbf{aCRN} increasing with $N_{max}$, suggesting that it continues to identify promising trajectories while other methods may be beginning to exhaust their search space. For $N_{max}=300$ we see that both \textbf{aCRN} and \textbf{fgCRN} outperform the hetGP-based methods, while for the larger two budgets, \textbf{fHet} and \textbf{aHet} surpass \textbf{fgCRN}. This indicates that the trajectory-oriented approach is successful early in the TS iterations, but as exploration continues, the adaptivity of the grid search allows for a richer exploration of the parameter space. 

\begin{figure}[!ht]
    \centering
    \includegraphics[width=\textwidth]{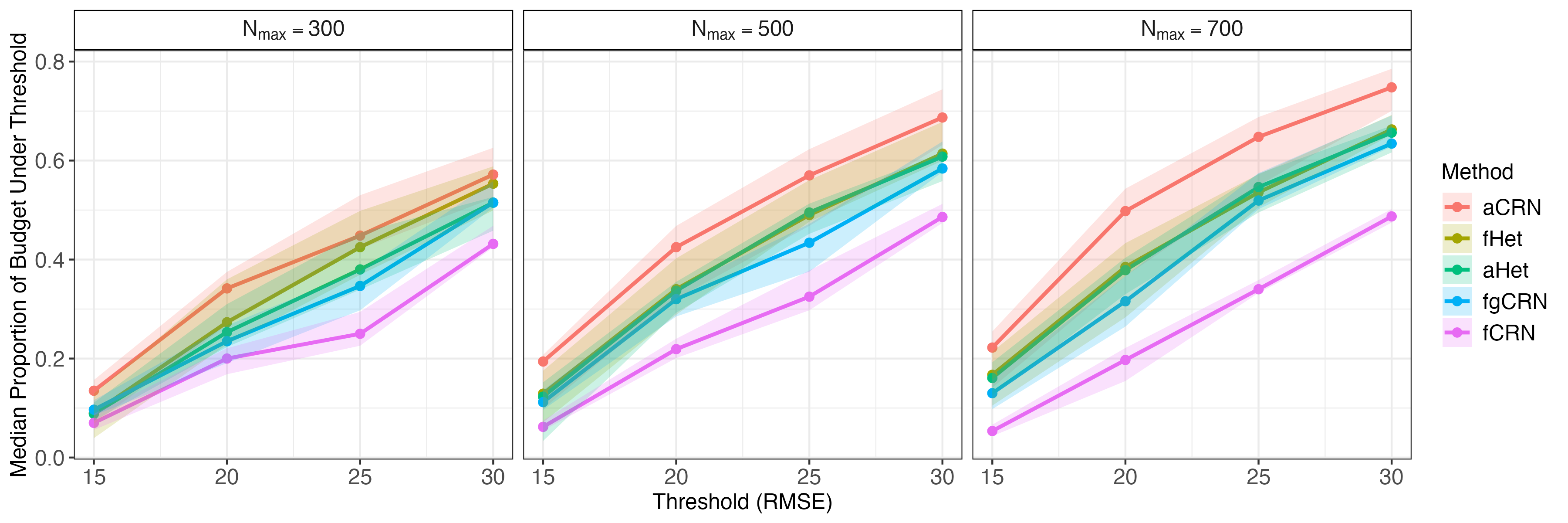}
    \caption{Means and standard errors of proportions of below-threshold trajectories found for the best-performing experimental design for each method.}
    \label{fig:pct_found_best}
\end{figure}

\begin{figure}[!ht]
    \centering
    \includegraphics[width=\textwidth]{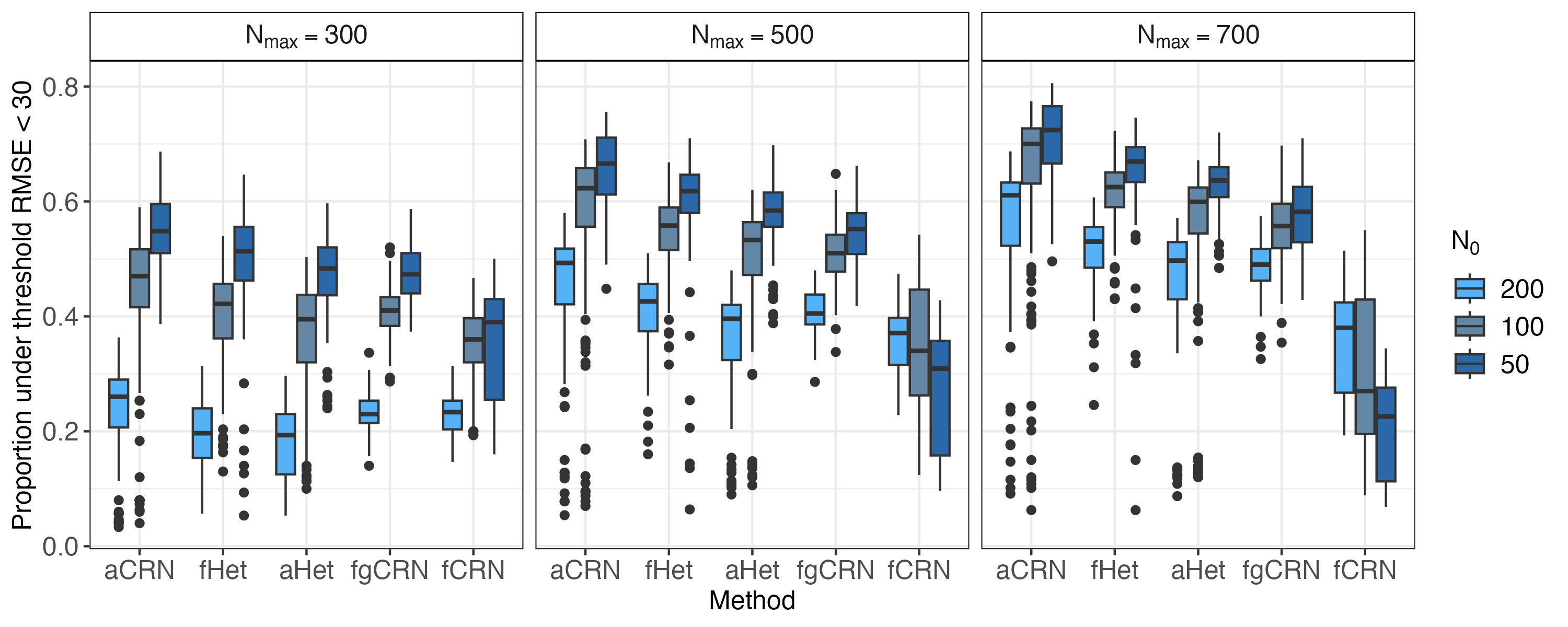}
    \caption{Proportion of trajectories with RMSE below 30 across all experiments varying the hyper-parameters, stratified by simulation budget $N_{\max}$, and initial design size $N_0$.}
    \label{fig:initial-design}
\end{figure}


We also compare results from the best-performing experimental design for each method.  For this comparison, we compute, for each method, the median proportion of below-threshold trajectories across the 10 experimental replicates associated with its optimal hyperparameter setting. As shown in~\Cref{fig:pct_found_best}, \textbf{aCRN} consistently finds a higher proportion of trajectories with RMSE below each threshold across all simulation budgets \(N_{\max} \in \{300, 500, 700\}\). The performance gap between aCRN and the comparator methods is most notable at stricter thresholds and higher $N_{\max}$, indicating that aCRN is more efficient in concentrating evaluations in regions of the (parameter, seed) space that yield high-quality trajectories.

Furthermore,~\Cref{fig:initial-design} shows that all algorithms, being sequential in nature benefit from smaller initial designs, that is, allocating a larger fraction of the total simulation budget to adaptive sampling rather than to the initial space-filling design. This enables the acquisition mechanism to more efficiently sample the (parameter, seed) space over informative regions. Here too, we observe that \textbf{aCRN} outperforms the other methods in collecting more high-quality trajectories across different simulation budget and initial design size.   


\subsubsection{Fast Time-to-Solution}
In addition to assessing the overall quality of the trajectories each method identifies within a given simulation budget, we want to assess the \textit{time-to-solution} of each method, that is, how \textit{quickly} high-quality trajectories are found through the optimization procedure. This metric is particularly relevant in epidemiology, where simulation runs can be expensive and timely inference is essential for updating forecasts or informing real-time decision-making. Methods that identify promising trajectories earlier in the search process allow resources to be allocated more efficiently and reduce the turnaround time required to obtain actionable outputs.

\begin{figure}[!t]
    \centering
    \includegraphics[width=.8\textwidth]{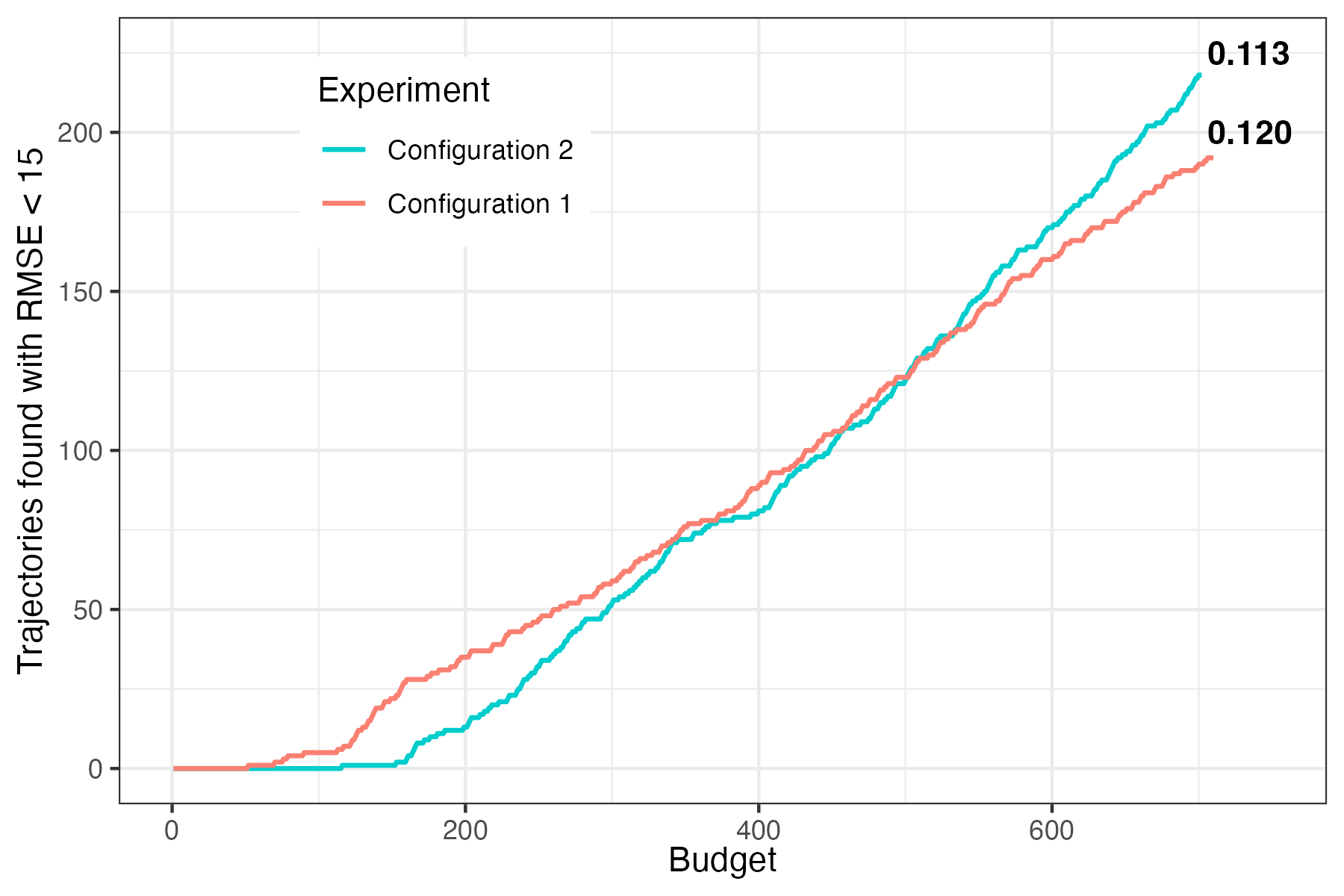}
    \caption{Number of trajectories with RSME $<$ 15 found using \textbf{aCRN} and \textbf{fHet} for two different experiments: Configuration 1 = $(n_{init}: 10, n_{rep}: 10, M: 300, n_{TS}: 10, \text{method: aCRN})$, Configuration 2 = $(n_{init}: 5, n_{rep}: 10, M: 300, n_{TS}: 20, \text{method: fHet})$. The rAUC values for each experiment are shown at the end of the search pathways.}
    \label{fig:auc_v_traj}
\end{figure}

\begin{figure}[!ht]
    \centering
    \includegraphics[width=\textwidth]{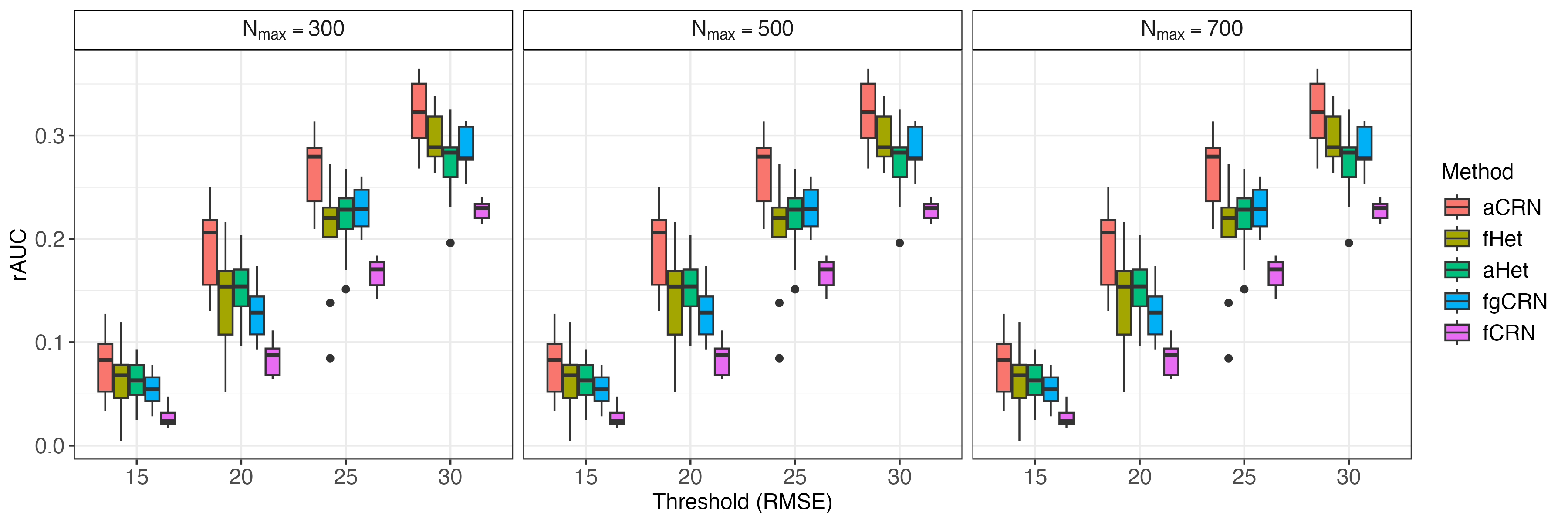}
    \caption{rAUC of best performing design for each method under different budget and varying thresholds.}
    \label{fig:auc_comparison}
\end{figure}

The time-to-solution is measured by counting the number of trajectories with RMSE below the specified threshold per budget expenditure. To quantify this metric across experiments, we estimate the area under the curve (AUC) of each search pathway using the standard trapezoidal formulation, and report a relative AUC (rAUC), normalized by $(N_{max})^2$. Specifically, letting $QT_t$ be the number of quality trajectories found by time $t$, we define
\begin{equation}\label{eq:rAUC}
\text{rAUC} := \frac{1}{(N_{max})^2}\sum_{t=2}^{N_{max}}\frac{(QT_t + QT_{t-1})}{2},    
\end{equation}
where higher values of rAUC correspond to methods that identify high-quality trajectories earlier and more consistently over time. This metric highlights temporal differences between methods that may not be captured by end-of-budget performance alone.
\Cref{fig:auc_v_traj} illustrates two representative search pathways from different experiments using \textbf{aCRN} (configuration 1) and \textbf{fHet} (configuration 2). Although configuration 1 ends up with lower number of below-threshold trajectories at the end of full simulation budget of 700, it discovers high-quality trajectories much earlier in the search process, achieving a higher rAUC. This emphasizes that end-of-budget summaries may obscure meaningful differences in search efficiency.
In~\Cref{fig:auc_comparison} we show the distribution of this metric across all experiments, where we again observe that \textbf{aCRN} consistently achieves the highest rAUC across simulation budgets and thresholds. The results suggest that \textbf{aCRN} is not only able to find more high-quality trajectories, but also does so at a faster rate compared to other methods. We also include additional experiments in the supplementary materials to demonstrate the robustness of the results.

\subsubsection{Parameter Exploration}
\begin{figure}[!h]
    \centering
    \includegraphics[width=1\textwidth]{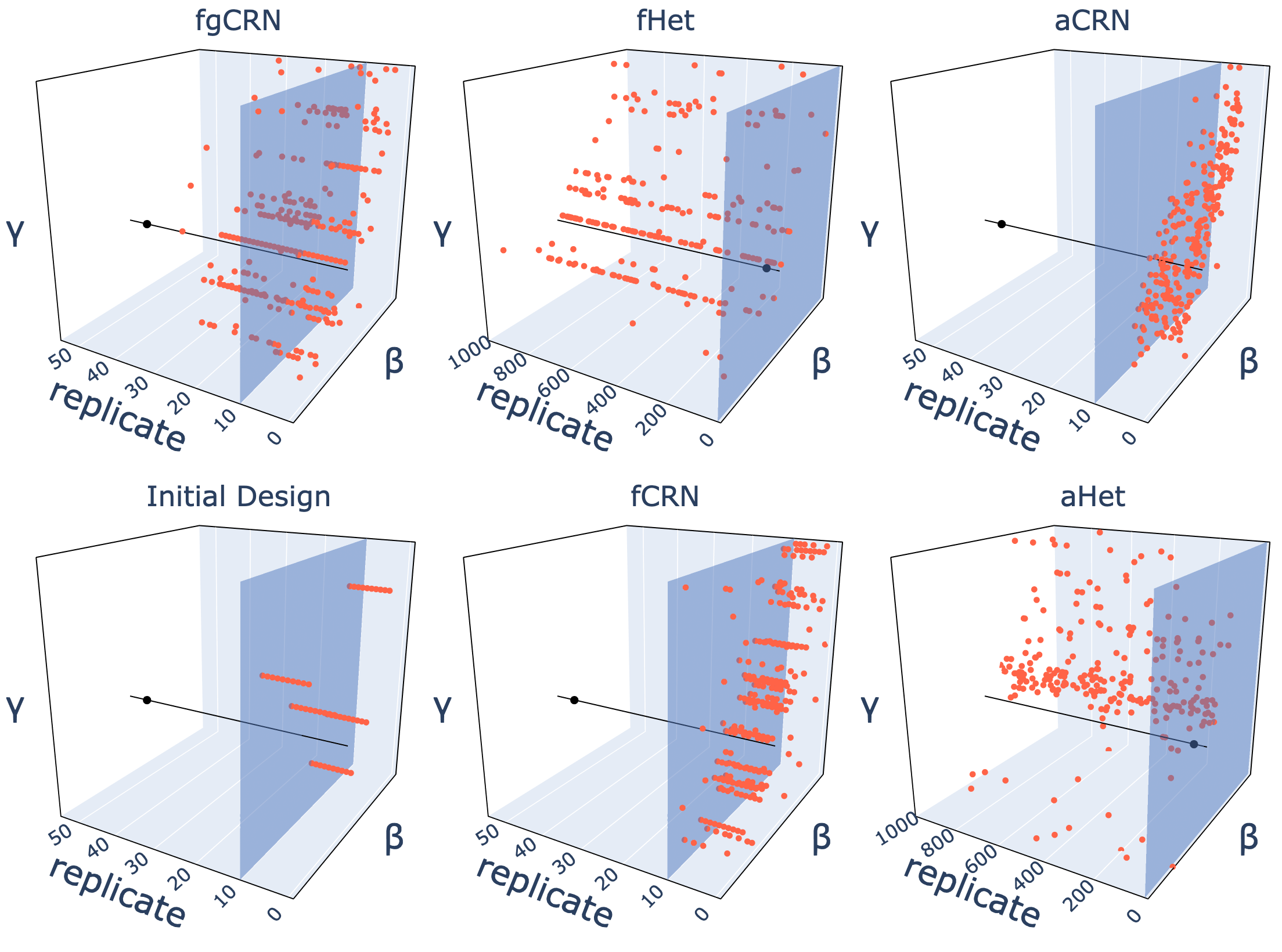}
    \caption{Sampling behavior of the comparator methods along with the initial design, across the joint model parameter space $(\beta, \gamma)$ and replicates. The red circles show the samples from each of the methods. The black dots each each panel indicate the location of the ground-truth parameter setting used in the experiments.}
    \label{fig:3d}
\end{figure}

\begin{figure}[!ht]
    \centering
    \includegraphics[width=\textwidth]{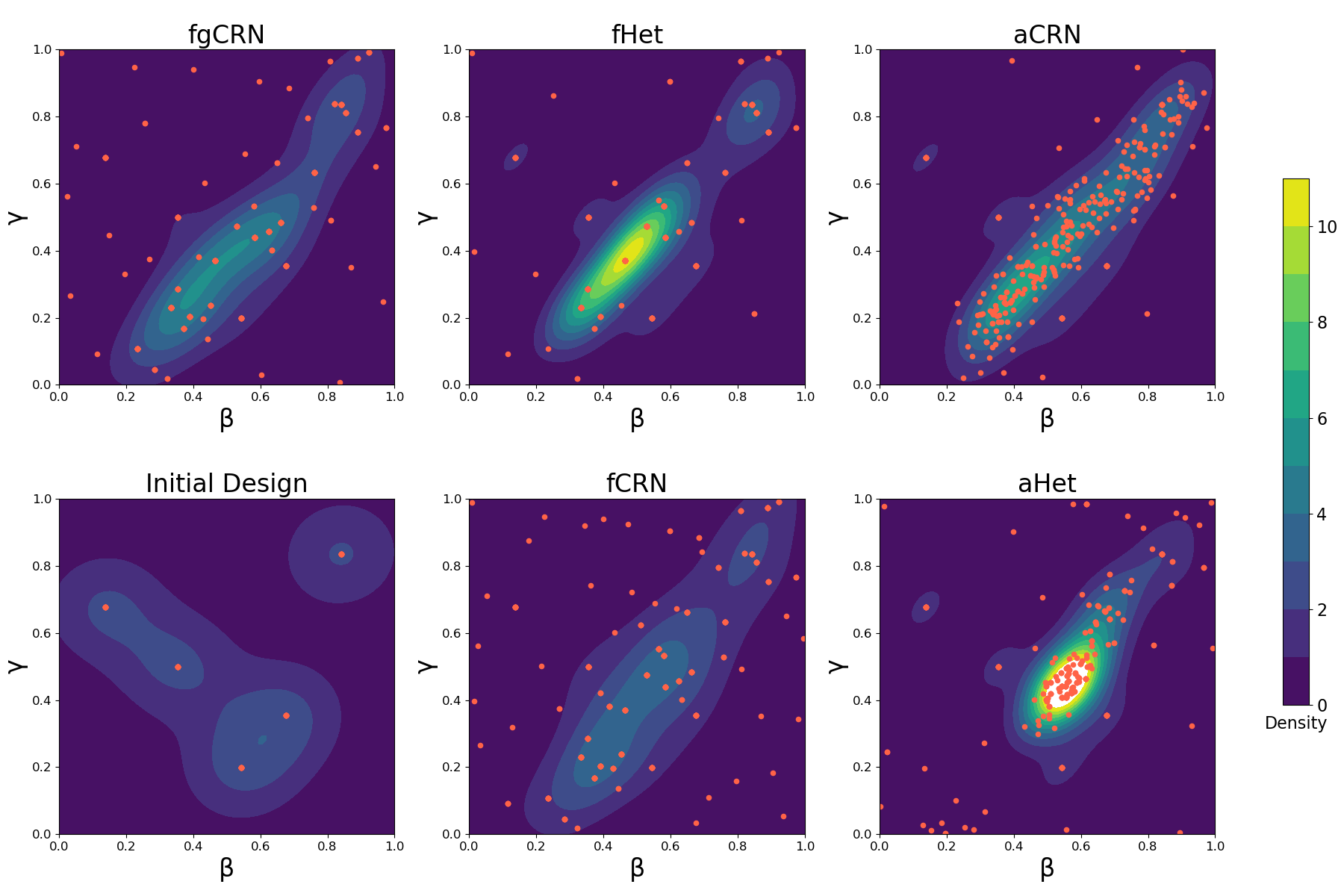}
    \caption{Sampling behavior of all the methods along with the initial design, across the two dimensional model parameter $(\beta, \gamma)$, and the density of samples in the background.}
    \label{fig:contour_all_6}
\end{figure}

In the case of a stochastic epidemiological model, such as our simple SIR model, or a more complex one such as CityCOVID, calibration parameters typically interact with underlying model stochasticity, such as population mixing patterns. Since each random seed induces a different mixing pattern for the present SIR model, the best set of calibration parameters may differ from one seed to another. For example, a random seed leading to more contacts than another may identify a lower $\beta$ in the former and higher in the latter, since the infection rate models both inherent pathogen transmissibility as well as population contact structure. Thus, it is of interest to study whether a TOO approach can facilitate increased exploration of parameter space to identify suitable trajectories. \Cref{fig:3d} compares the sampling behavior of the six methods across the joint model parameter space $(\beta, \gamma)$ and replicate identifiers $r$. Each red dot represents an evaluated input-replicate pair $(\xb, r)$, and the shaded plane indicates the restricted replicate set used for each. \Cref{fig:contour_all_6} shows the densities of the samples projected to the ($\beta$,$\gamma$) plane.
The black dot denotes the ground-truth parameter setting used to generate the observed data.

We observe that \textbf{fHet} exhibits broad sampling along the replicate axis (\Cref{fig:3d})) but limited diversity across the parameter space (\Cref{fig:contour_all_6})). In contrast, \textbf{aCRN}, which explicitly includes the replicates in its search space, achieves broader parameter coverage (\Cref{fig:3d}c) and d)).


\section{Finding Trajectories in a City-scale Agent-based Model}
\label{sec:usecase}

The stochastic SIR model in the preceding section provides a convenient testbed for evaluating optimization strategies, but it is too simplistic to capture many of the complexities of real-world epidemics. Here we illustrate the performance of our proposed method on a more realistic and computationally expensive model by applying it to \textit{CityCOVID}~\cite{Ozik2021, macal2020citycovid}, a high-resolution ABM developed during the COVID-19 pandemic to support public health decision making in the Chicago metropolitan area.

\subsection{CityCOVID Simulation Model}

CityCOVID simulates the behavior and interactions of approximately 2.7 million agents, each representing an individual in Chicago, with assigned daily activity schedules that include commuting, school, work, and other location-based contacts. Disease transmission is modeled as a stochastic process, and key epidemiological outcomes (e.g., infections, hospitalizations, deaths) are output over time. Stochasticity arises not only from the transmission process itself but also from initial conditions, including the seeding of infections, and individual movement patterns. This suggests that a more deliberate treatment of stochasticity, beyond the use of marginal summaries, could be beneficial.

Historically, calibrating such a complex model has posed considerable challenges. The initial calibration of CityCOVID used incremental mixture approximate Bayesian computation (IMABC)~\cite{rutter2019microsimulation}, a likelihood-free method that required approximately 35,000 simulations to arrive at a plausible posterior distribution. While recent efforts have attempted to reduce the computational burden and improve uncertainty quantification (e.g., via surrogate models~\cite{Ozik2021, binoisPortfolioApproachMassively2025, robertson2025bayesian}), these approaches have focused on matching the mean response of the simulator to observed data. Such strategies may obscure important trajectory-level variations and fail to fully account for the structural role of randomness in shaping epidemic dynamics. As shown in~\Cref{fig:citycovid_top50}, even parameter configurations calibrated to the mean response can yield a wide range of individual trajectories, many of which deviate significantly from the observed data, while their mean behavior is consistent with observed data. 

Initial CityCOVID calibration efforts \cite{Ozik2021} identified nine relevant parameters via a Morris global sensitivity analysis \cite{morris1991}, which was shown to be further reducible to four parameters in \cite{robertson2025bayesian}. Here we further constrain one of these parameters, the time of initial exposure to infection, to a fixed value and focus on three parameters (\Cref{table:citycovid_priors}). 


\begin{table}[h!]
\centering
\label{table:citycovid_priors}
\begin{tabular}{l l l l} 
 \toprule
 \textbf{Parameter} & \textbf{Description} & \textbf{Prior} \\ [0.5ex] 
 \midrule
$\theta_1$ & Susceptible to exposed probability & $U(0.01, 0.15)$\\ 
$\theta_2$  & Stay at home probability & $U(0.1, 1)$ \\
$\theta_3$  & Behavioral adjustment probability & $U(0.01, 0.5)$ \\ 
[1ex] 
 \bottomrule
\end{tabular}
\vspace{0.2cm}
\caption{CityCOVID parameters and their priors used for trajectory finding}
\end{table}

\subsection{Calibrating to Both Hospitalizations and Deaths}

As in previous work, we compare model outputs to the two empirical datasets that were the most reliably reported and available during the COVID-19 pandemic: the daily count of hospitalizations and deaths attributed to COVID-19 from April through June 2020, which was accessed through the Chicago Data Portal~\cite{CityOfChicago2021}. Let $h_t$ and $d_t$ be the empirical counts of deaths and hospitalizations, respectively, at time $t$. Since we are concerned not just with parameters $\boldsymbol{\theta}$ but also the random seed $r$, let $\boldsymbol{\theta}_r := (\theta_1, \theta_2, \theta_3, r)$ specify inputs that generate a particular CityCOVID trajectory. Let $h_{\boldsymbol{\theta}_r,t}$ and $d_{\boldsymbol{\theta}_r,t}$ denote the simulated count of hospitalizations and deaths. We define the dual-objective function
\begin{equation}
\mathcal{L}(\boldsymbol{\theta}_r) := \sum_{t=1}^T\frac{|h_t - h_{\boldsymbol{\theta}_r,t}|}{h_t} + \sum_{t=1}^T\frac{|d_t - d_{\boldsymbol{\theta}_r,t}|}{d_t},    
\end{equation}
\noindent where $t=1, \cdots,  T$ denotes the time indices where observations are made (typically a day). This dual-objective function helps ensure the trajectories we identify replicate the epidemic pathway across multiple outcomes.

\subsection{Experiments}

For these experiments, we set our simulation budget to $N_{max}=3000$ and compare our $\textbf{aCRN}$ method to the next best performing method, $\textbf{fHet}$. 
We initialize the experiments with a space-filing design of size $n_{init} = 20$ over $(\theta_1, \theta_2, \theta_3)$ following the prior distributions specified in \Cref{table:citycovid_priors}, and replicate each parameter setting $n_{rep}=30$ times, for a total initial budget of $N_0 = 600$. The remaining hyper-parameters (\Cref{table:exp_params}) are set to $M=100$ and $n_{TS}=100$.

These experiments were performed on Argonne Laboratory Computing Resource Center's Improv cluster. We relied on the ability of EMEWS to enable complex algorithms, such as $\textbf{aCRN}$ and $\textbf{fHet}$, to direct the iterating logic of large HPC workflows, in this case coordinating the running of large ensembles of MPI-distributed simulations. Additional details can be found in the Supplementary Materials.

\subsection{Results}

\begin{figure}[!t]
    \centering
    \includegraphics[width = \textwidth]{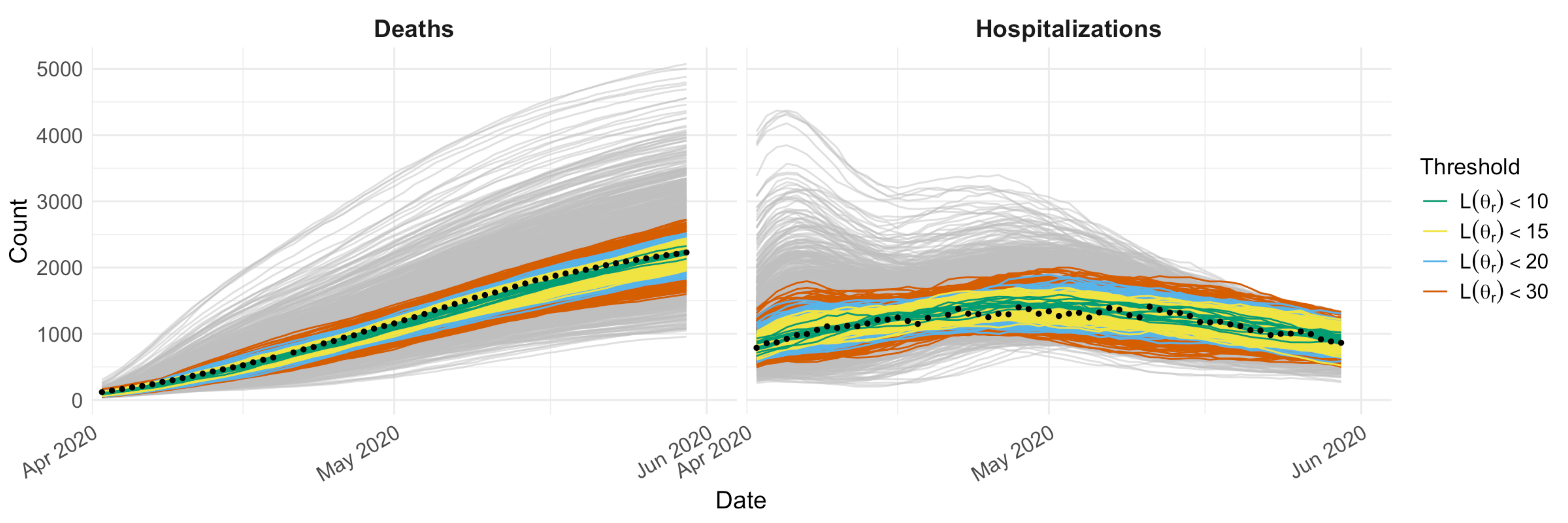}
    \caption{Trajectories of deaths (left) and hospitalizations (right) from CityCOVID using aCRN for varying thresholds denoted by colored lines, and observed data (black dots). }
    \label{fig:citycovid_imabc_thresh}
\end{figure}

After exhausting the simulation budget for each method, we assess the quality of the trajectories that were found. Similar to~\Cref{sec:simulation_results}, trajectory quality is assessed relative to thresholded values, in this case of the dual-objective function $\mathcal{L}(\boldsymbol{\theta}_r)$. For illustrative purposes in~\Cref{fig:citycovid_imabc_thresh} we show the trajectories found by $\textbf{aCRN}$, color-coded by their threshold values.

\Cref{table:citycovid_th_results} shows that at the end of the simulation budget, \textbf{aCRN} does a better job than \textbf{fHet} at identifying high-quality trajectories across most $\mathcal{L}(\boldsymbol{\theta}_r)$ thresholds. For the strictest criteria (i.e., $\mathcal{L}(\theta_r) < 10$ and $\mathcal{L}(\theta_r) < 15$), which represent trajectories most closely matching deaths and hospitalizations data, \textbf{aCRN} substantially overperforms. As the threshold is relaxed, the difference between the two methods narrows, ultimately resulting in a higher number of trajectories found by \textbf{fHet} when $\mathcal{L}(\theta_r) < 30$. 

\begin{table}[h!]
\centering
\begin{tabular}{l c c c c} 
 \hline
& $\mathcal{L}(\boldsymbol{\theta}_r) < 10$ & $\mathcal{L}(\boldsymbol{\theta}_r) < 15$ &$\mathcal{L}(\boldsymbol{\theta}_r) < 20$ &$\mathcal{L}(\boldsymbol{\theta}_r) < 30$  \\ [0.5ex] 
 \hline
$\textbf{aCRN}$ & 4 & 52 & 149 & 698\\ 
$\textbf{fHet}$  & 1 & 1 & 53 & 923\\
 \hline
\end{tabular}
\vspace{0.2cm}
\caption{Number of trajectories found by each method with objective function lower than stated thresholds. Counts are relative to a total simulation budget of 3000.}
\label{table:citycovid_th_results}
\end{table}

\begin{figure}[!ht]
    \centering
    \includegraphics[width = \textwidth]{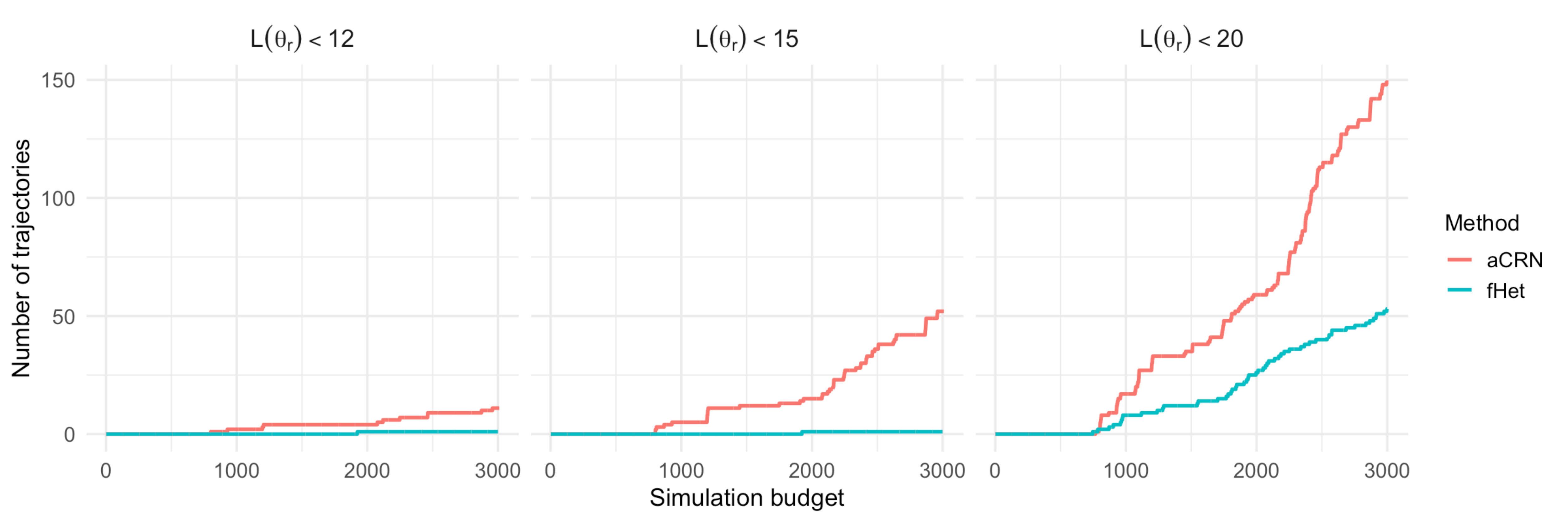}
    \caption{Number of trajectories found per simulation budget expenditure with objective function below the specified threshold.}
    \label{fig:citycovid_traj_found}
\end{figure}

From a fast-time-to-solution perspective,~\Cref{fig:citycovid_traj_found} demonstrates that  \textbf{aCRN} begins identifying high-quality trajectories early, quickly accumulating them as the simulation budget is increased. This is particularly important when expensive-to-run simulations, such as the CityCOVID model used here, are used for decision support purposes. 

Further examination of the regions of parameter space that are explored for the $\mathcal{L}(\theta_r) < 15$ threshold shows that \textbf{aCRN} finds trajectories across a much more diverse region, capturing multiple plausible solutions, whereas \textbf{fHet} concentrates almost exclusively on a single grid point in the fixed grid, repeatedly sampling additional replicates around it. This highlights the advantage of combining the CRNGP surrogate with adaptive-grid refinement, allowing \textbf{aCRN} to efficiently balance exploration of new parameter regions with exploitation of high-likelihood areas (see Figure~S16 in the Supplementary Materials).

\begin{figure}[!h]
    \centering
    \includegraphics[width = 0.8\textwidth]{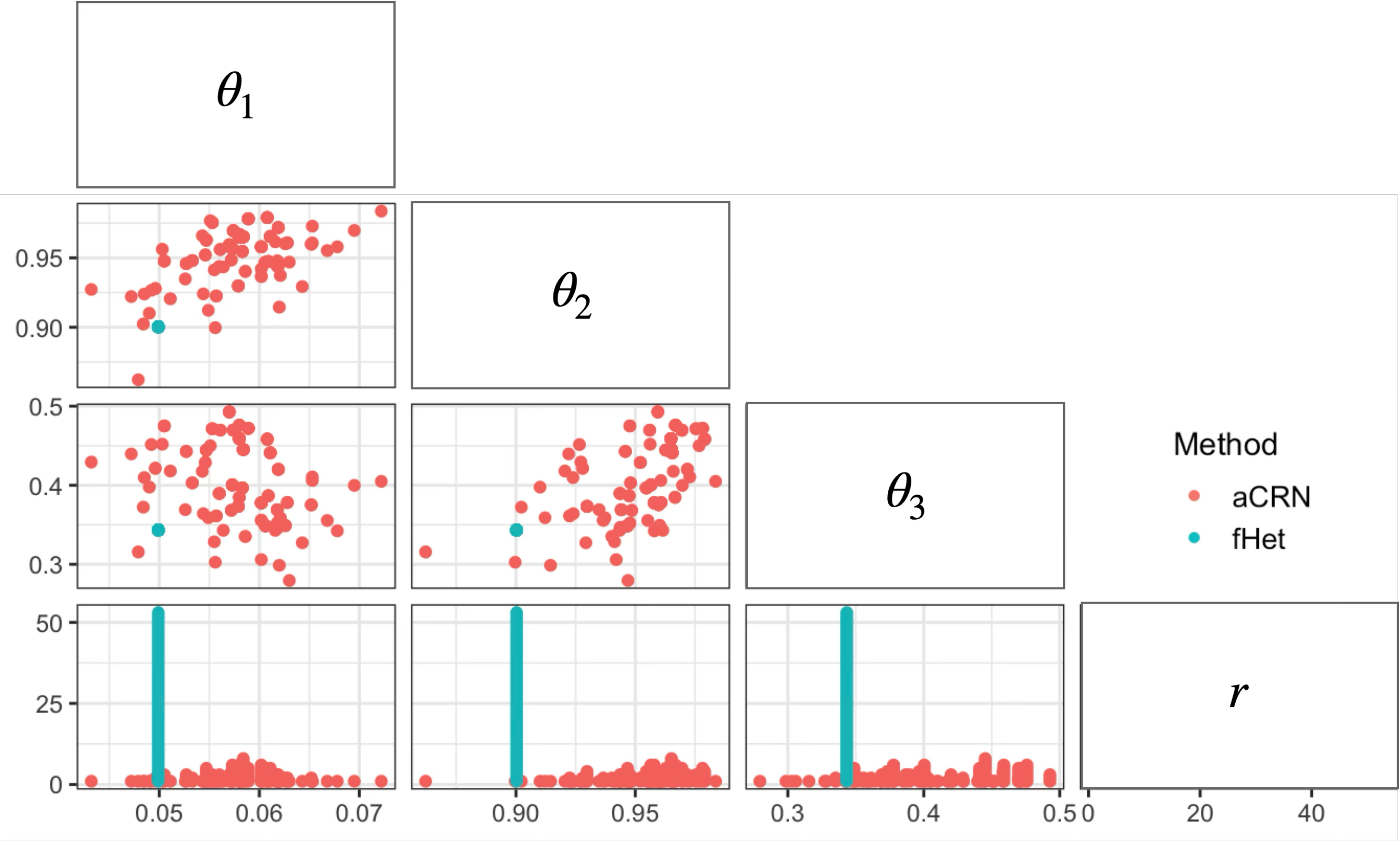}
    \caption{Parameter-replicate tuples corresponding to the trajectories with $\mathcal{L}(\boldsymbol{\theta}_r) < 15$ found by \textbf{aCRN} and \textbf{fHet}.}
    \label{fig:citycovid_pairs}
\end{figure}



\section{Conclusion and Future Work}\label{sec:conclusion}



We proposed an adaptive Bayesian optimization framework built around the CRNGP surrogate and Thompson Sampling, designed to efficiently discover parameter–seed pairs that align closely with observed data. The CRNGP surrogate enables realization-level prediction by modeling simulation outputs as functions of both input parameters and replicate identifiers. To improve sampling efficiency and improve time-to-solution, we introduced an adaptive-grid strategy that iteratively filters and densifies candidate input locations, thereby concentrating computational effort on regions of high posterior utility. One of the primary motivators of the work is the desire to use complex simulations to produce actionable insights for decision makers within decision-relevant timelines. Furthermore, for stochastic epidemic models, random seeds control specific instantiations of population mixing and epidemic evolution. Thus, identifying individual trajectories, rather than only model parameters, allows for capturing empirically consistent snapshots, through model checkpoints, that can be used as initial conditions for subsequent analyses, such as data assimilation as new data is obtained or intervention optimization analyses.  

Several avenues for future work remain. First, strategies for selecting and updating the replicate set could be extended beyond the fixed design considered here. Systematic growth of the replicate set, guided by updated GP draws~\cite{chevalier2015fast} or their continuous approximations~\cite{mutny2018efficient}, would increase flexibility while maintaining computational tractability. Second, explicit model-discrepancy terms could be integrated into the optimization framework to account for differences between simulator outputs and observed data~\cite{koermer2024augmenting}. Third, the current CRNGP covariance function assumes a uniform similarity index across seeds (\Cref{eq:crncov}); more general formulations could leverage grouping structures for categorical inputs~\cite{roustant2020group} or exploit more complex relationships between seeds in latent spaces~\cite{da2025distributional}. Finally, while our framework focused on scalar summary outputs, extensions to functional or temporal outputs may yield further improvements, as demonstrated in related work on multi-output and functional BO~\cite{astudillo2021thinking}. Although motivated here by epidemic modeling, the proposed framework is general and directly applicable to a wide range of fields that employ stochastic simulators, where realization-level calibration may be beneficial for accurate inference and decision support.

\section*{Acknowledgments}
This research was completed with resources provided by the Argonne Computing, Environment, and Life Sciences Directorate General Computing Environment, the Laboratory Computing Resource Center at Argonne National Laboratory, and the Argonne Leadership Computing Facility, which is a DOE Office of Science User Facility. 

\section*{Funding}
This material is based upon work supported by the National Science Foundation under Grant 2200234, the U.S. Department of Energy, Office of Science, under contract number DE-AC02-06CH11357 and the Bio-preparedness Research Virtual Environment (BRaVE) initiative.




\bibliographystyle{plain} 
\bibliography{crngp_bib}       

\clearpage

\section{Supplementary Materials}

\subsection{Experiment Workflows}

The experiments using the compartmental SIR model (programmed in R) in Section 4 were performed on Argonne Laboratory Computing Resource Center's Improv cluster using an EMEWS~\citep{collierDistributedModelExploration2024} sweep workflow~\citep{emews-sweep}. 
The sweep workflow takes a CSV format file as input where each row in the CSV file is a set of algorithm hyperparameters, i.e., a unique combination from~Table 1. Each hyperparameter is applied to all of the comparator algorithms. These parameters are parsed using Swift-T's~\citep{wozniakSwiftLargeScaleApplication2013} embedded Python interpreter and passed to the SIR model for evaluation. The SIR model is called using R's Rscript executable via a bash script invoked by Swift-T. This process, from the parameter parsing to the bash script Rscript call, occurs in parallel at a scale configured by the workflow's hardware specification. In this case, 2 nodes were used, with a processes per node (PPN) count of 64, allowing for 126 (2 processes are used by EMEWS) total parallel model runs. Additionally, we set the number of threads available to each SIR model invocation to 2, fully utilizing the 128 (64 model runs, using 2 threads each) CPU cores available on each Improv node.

The experiments involving CityCOVID (Section 5) were also performed on Argonne Laboratory Computing Resource Center's Improv cluster using a decoupled EMEWS DB workflow~\citep{emews-db}. Here the $\textbf{aCRN}$ and $\textbf{fHet}$ algorithms submit parameters for evaluation to an EMEWS database, running on an Improv login node, through an EQSQL task queue. Once a batch of parameters has been submitted, the \textbf{aCRN} or \textbf{fHet} algorithms (programmed in R) periodically poll the queue for results. When the full batch of results has been retrieved, the algorithm can proceed to the next round. A separate worker pool, configured to run CityCOVID as a parallel Swift/T leaf function \citep{woz_swift_guide}, retrieves the parameters to evaluate from the database and executes a CityCOVID instance with those parameters as inputs. When a CityCOVID run finishes, the location of an output file from which the number of hospitalizations and deaths can be calculated is reported back to the database. The algorithm code retrieves the file path from the task queue, and computes the number of deaths and hospitalizations from it. The overall workflow is run using a SLURM submission script that submits a single node job for the R algorithm code. When this executes, it 1) starts the EMEWS database, if necessary; 2) creates the EMEWS task queue; and 3) submits a multi-node worker pool job that runs the EMEWS worker pool code. In these experiments, the worker pool was configured to run on 17 nodes with a PPN of 128. Each CityCOVID instance runs in approximately 50 seconds on 256 processes (2 nodes), allowing for 8 instances to be executed concurrently. The remaining processes are used by EMEWS to manage the workflow, e.g., for retrieving tasks and reporting results.

\subsection{Results from Additional Simulated Experiments}

In addition to the main experiments reported in the paper, we performed supplementary experiments (A1, A2) to evaluate the robustness of the proposed methods across different choices of ground truth parameters. The experimental setup, including the optimization budget and evaluation metrics, is identical to that described in the main text; the only change is the parameter values used to generate the synthetic observed data. The specific ground truth values for each experiment are provided below.

In Experiment A1, the results are consistent with those reported in the main paper. I.e., that the trajectory-oriented methods, in particular \textbf{aCRN}, identify high-quality parameter–trajectory pairs more efficiently than approaches relying only on parameter-level inference. Figures S4–S8 summarize these findings. In contrast, Experiment A2 shows little distinction among methods, with no single approach demonstrating clear superiority. This outcome is explained by the ground truth used in A2, which generated very low infection counts (Figure S9), leading to limited signal in the data and consequently less pronounced differences across methods.

\subsubsection{Experiment A1}

The ground truth is generated by setting $\beta = 0.7$, $\gamma = 0.2$, and seed = 50.

\begin{figure}[!ht]
    \centering
    \includegraphics[width=0.7\textwidth]{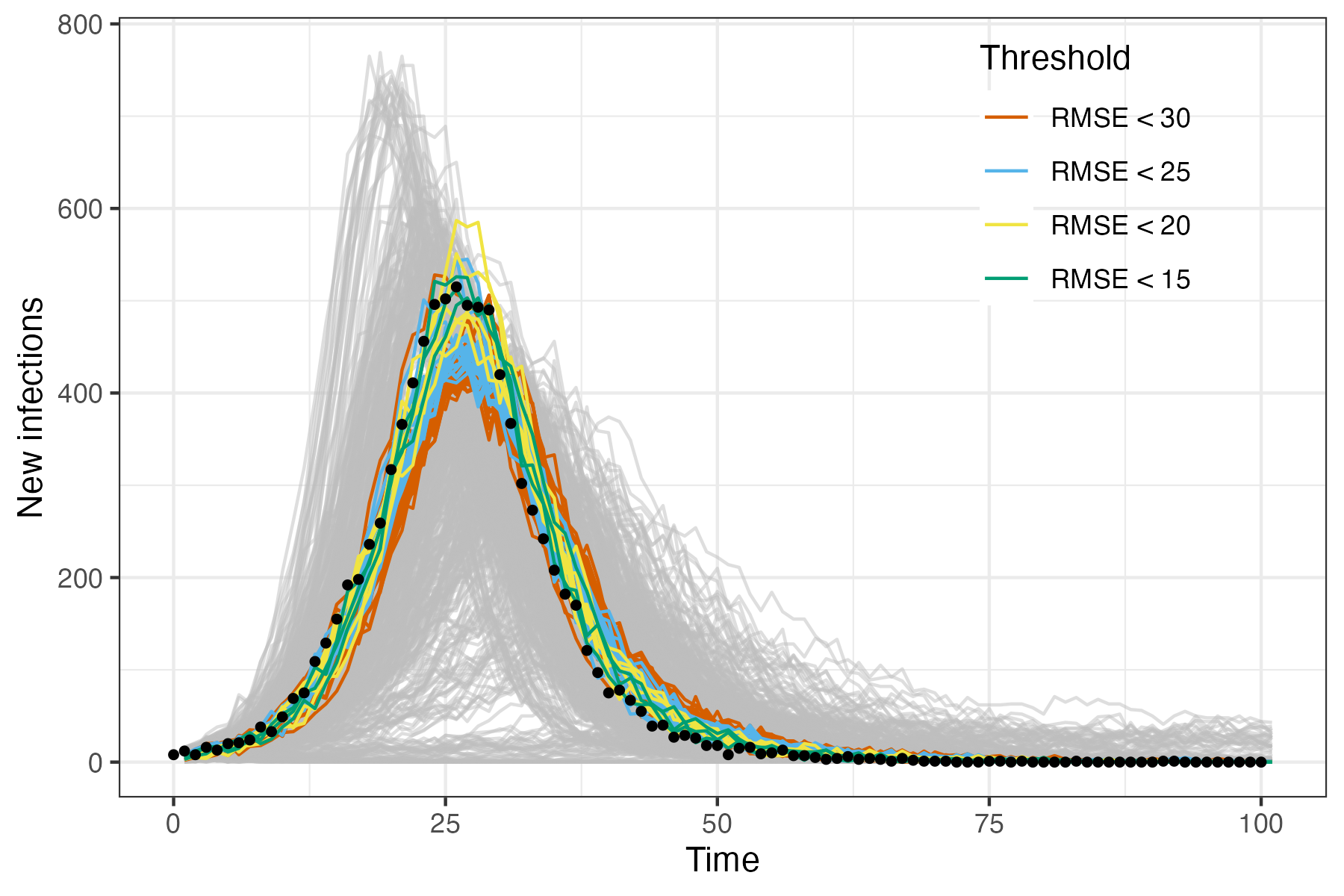}
    \caption{Simulated epidemic trajectories found by the aCRN method for a single experiment configuration. Each trajectory is colored according to its RMSE category, with thresholds at 15, 20, 25, and 30. The ground truth is shown using black dots.}
    \label{fig:thresholds_0.7_0.2}
\end{figure}

\begin{figure}[!ht]
    \centering
    \includegraphics[width=\textwidth]{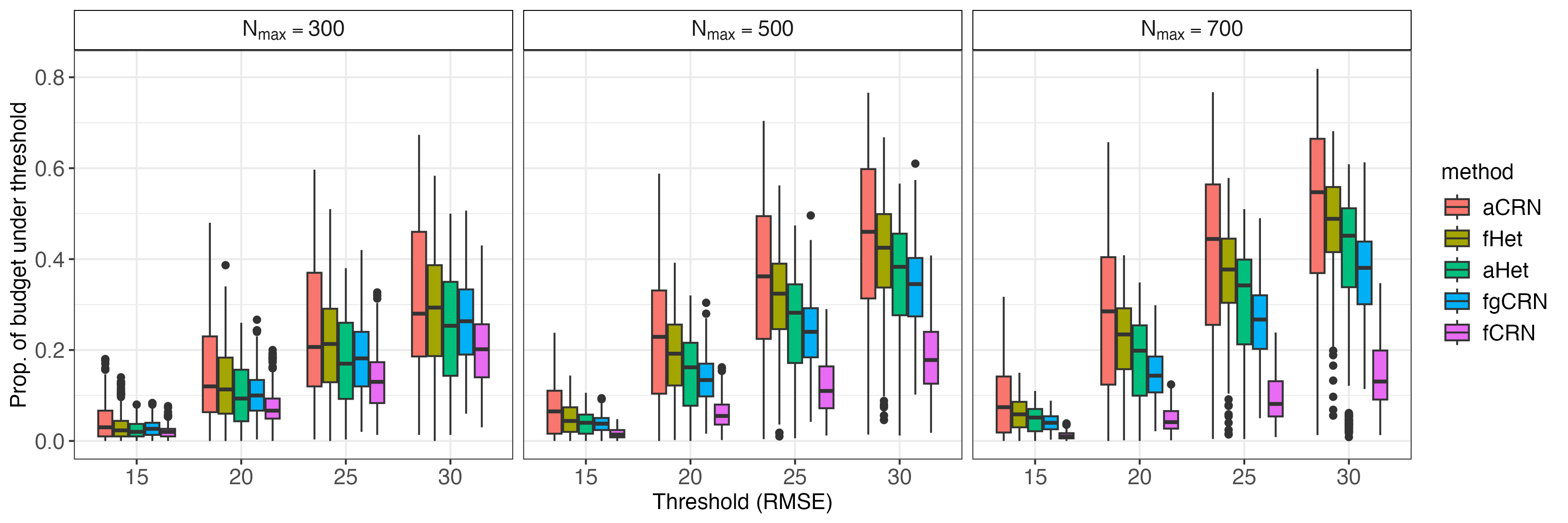}
    \caption{Proportion of trajectories with RMSE below varying thresholds across all experiments varying the hyper-parameters stratified by simulation budget $N_{\max}$.}
    \label{fig:overall_0.7_0.2}
\end{figure}

\begin{figure}[!ht]
    \centering
    \includegraphics[width=\textwidth]{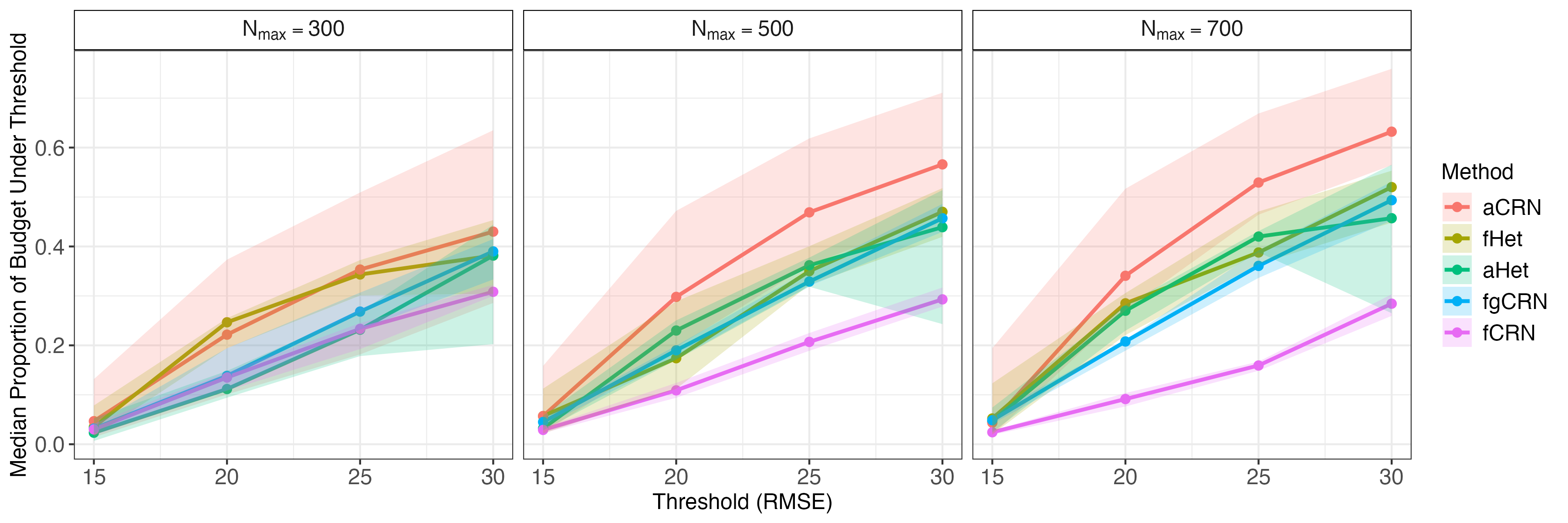}
    \caption{Means and standard errors of proportions of below-threshold trajectories found for the best-performing experimental design for each method.}
    \label{fig:pct_found_best_0.7_0.2}
\end{figure}

\begin{figure}[!ht]
    \centering
    \includegraphics[width=\textwidth]{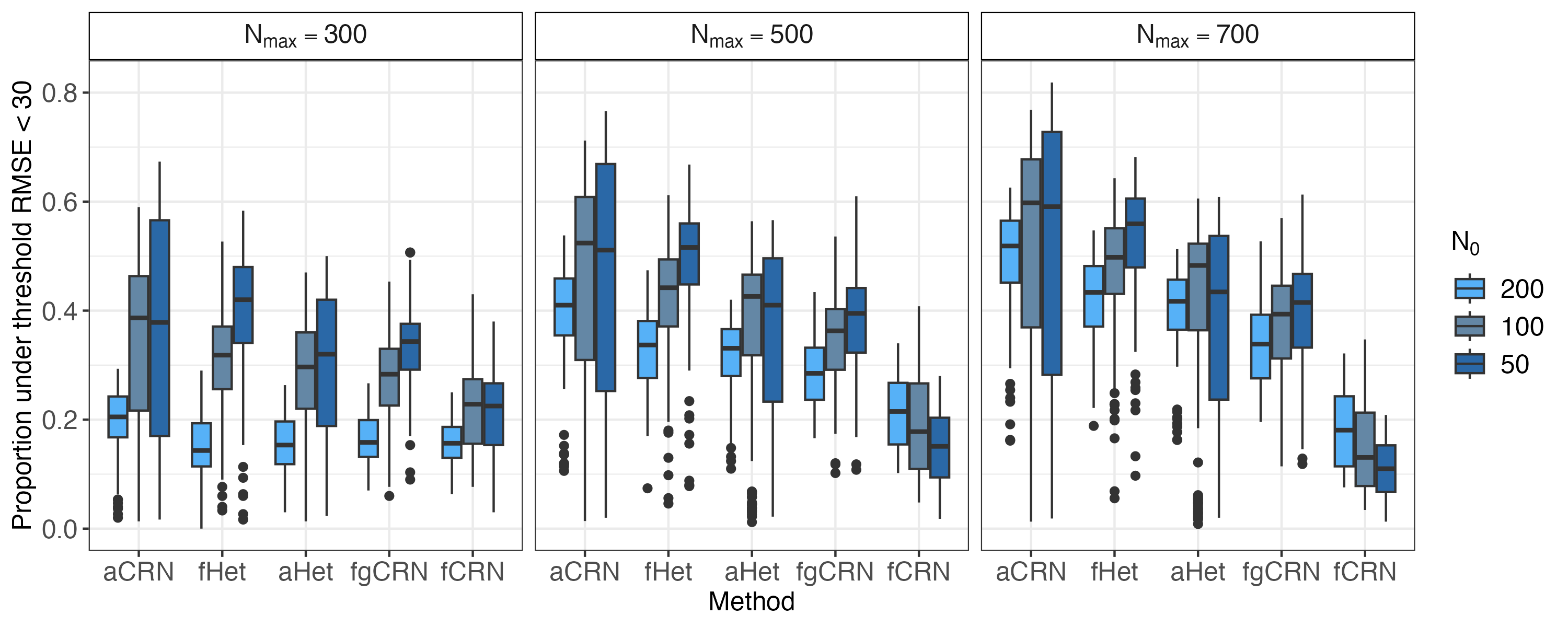}
    \caption{Proportion of trajectories with RMSE below 30 across all experiments varying the hyper-parameters, stratified by simulation budget $N_{\max}$, and initial design size $N_0$.}
    \label{fig:initial-design_0.7_0.2}
\end{figure}

\begin{figure}[!ht]
    \centering
    \includegraphics[width=\textwidth]{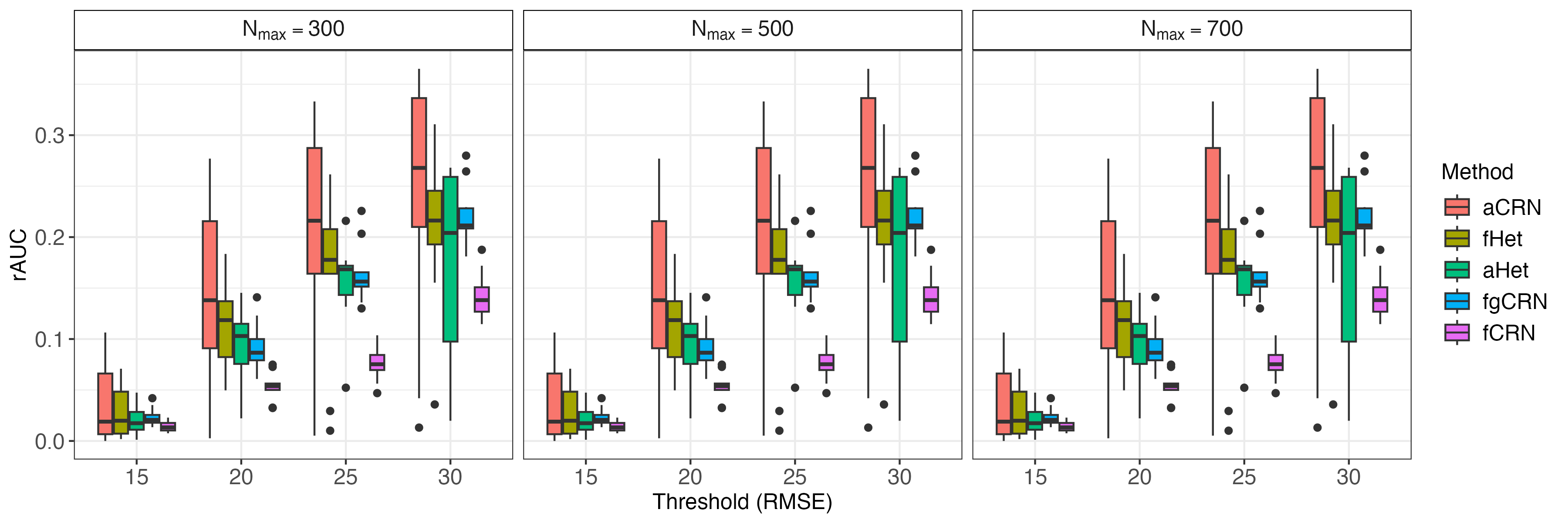}
    \caption{rAUC of best performing design for each method under different budget and varying thresholds.}
    \label{fig:auc_comparison_0.7_0.2}
\end{figure}

\clearpage

\subsubsection{Experiment A2}

The ground truth is generated by setting $\beta = 0.3$, $\gamma = 0.8$, and seed = 50.

\begin{figure}[!ht]
    \centering
    \includegraphics[width=0.7\textwidth]{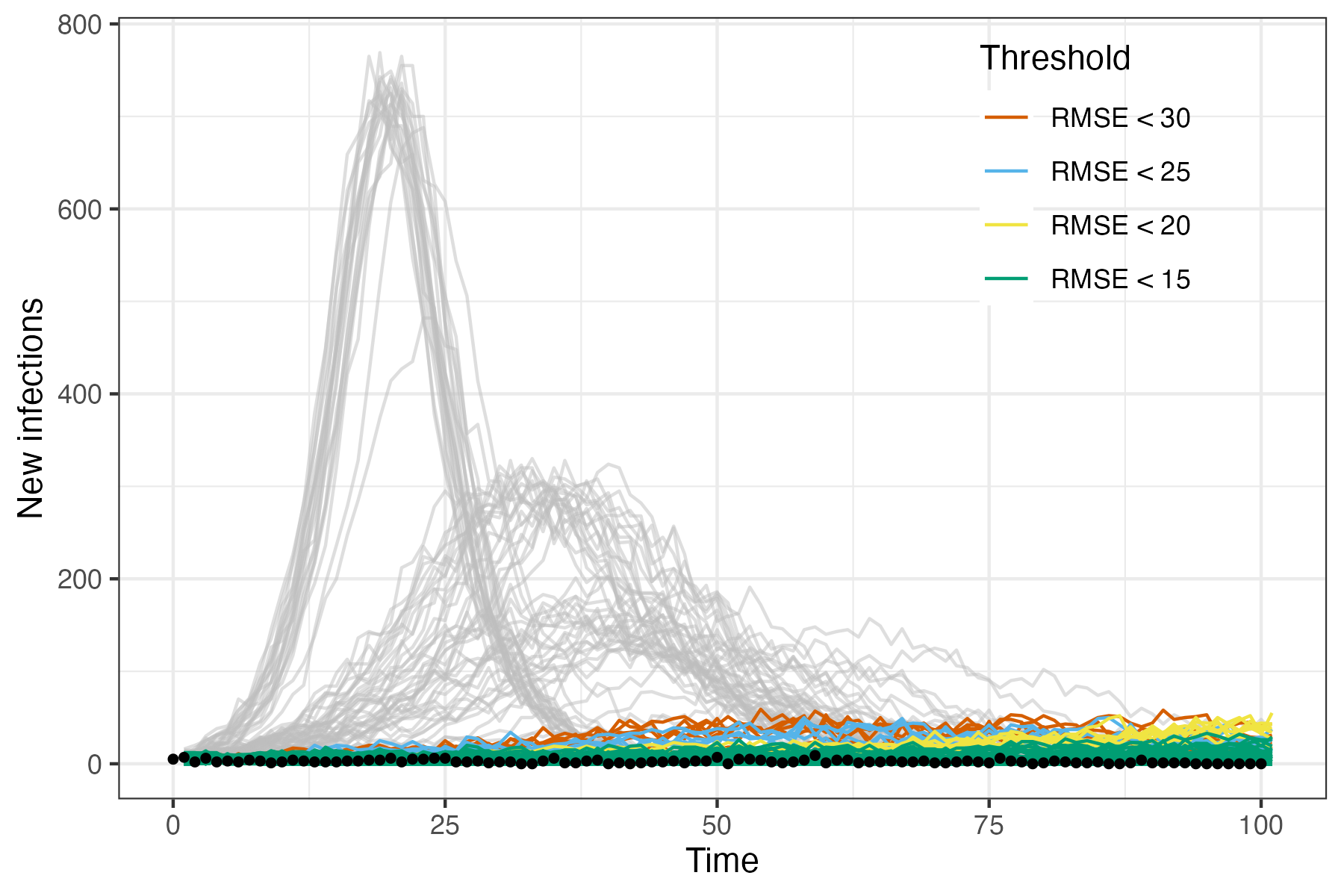}
    \caption{Simulated epidemic trajectories found by the aCRN method for a single experiment configuration. Each trajectory is colored according to its RMSE category, with thresholds at 15, 20, 25, and 30. The ground truth is shown using black dots.}
    \label{fig:thresholds_0.3_0.8}
\end{figure}

\begin{figure}[!ht]
    \centering
    \includegraphics[width=\textwidth]{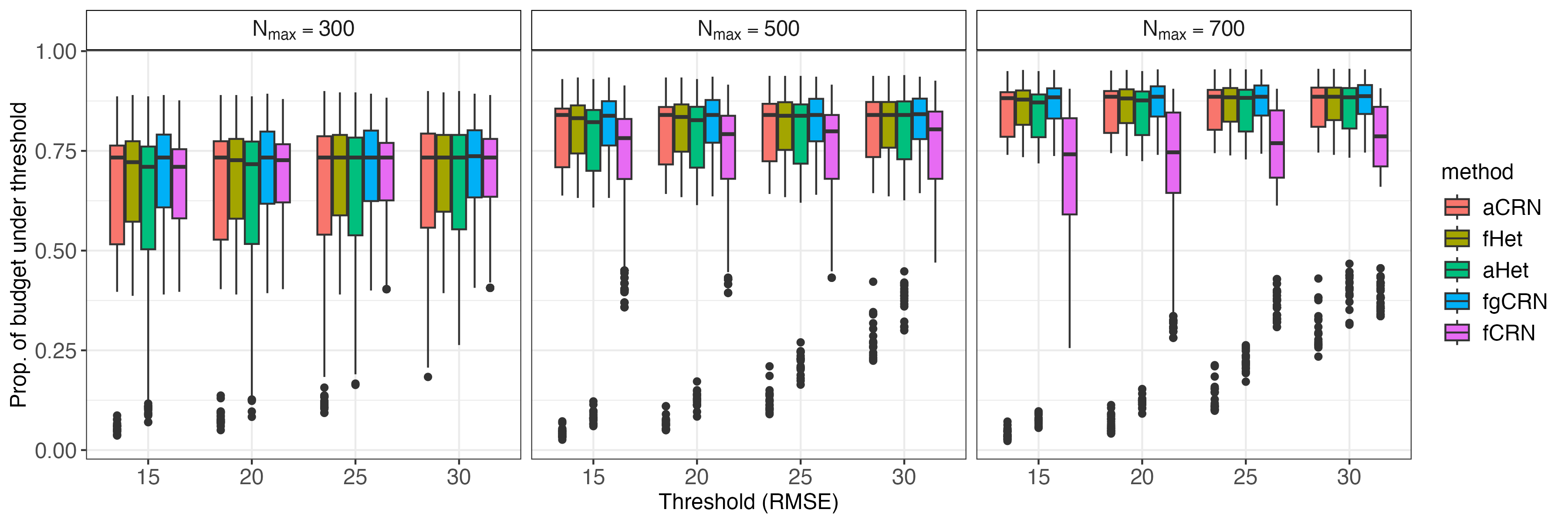}
    \caption{Proportion of trajectories with RMSE below varying thresholds across all experiments varying the hyper-parametersstratified by simulation budget $N_{\max}$.}
    \label{fig:overall_0.3_0.8}
\end{figure}

\begin{figure}[!ht]
    \centering
    \includegraphics[width=\textwidth]{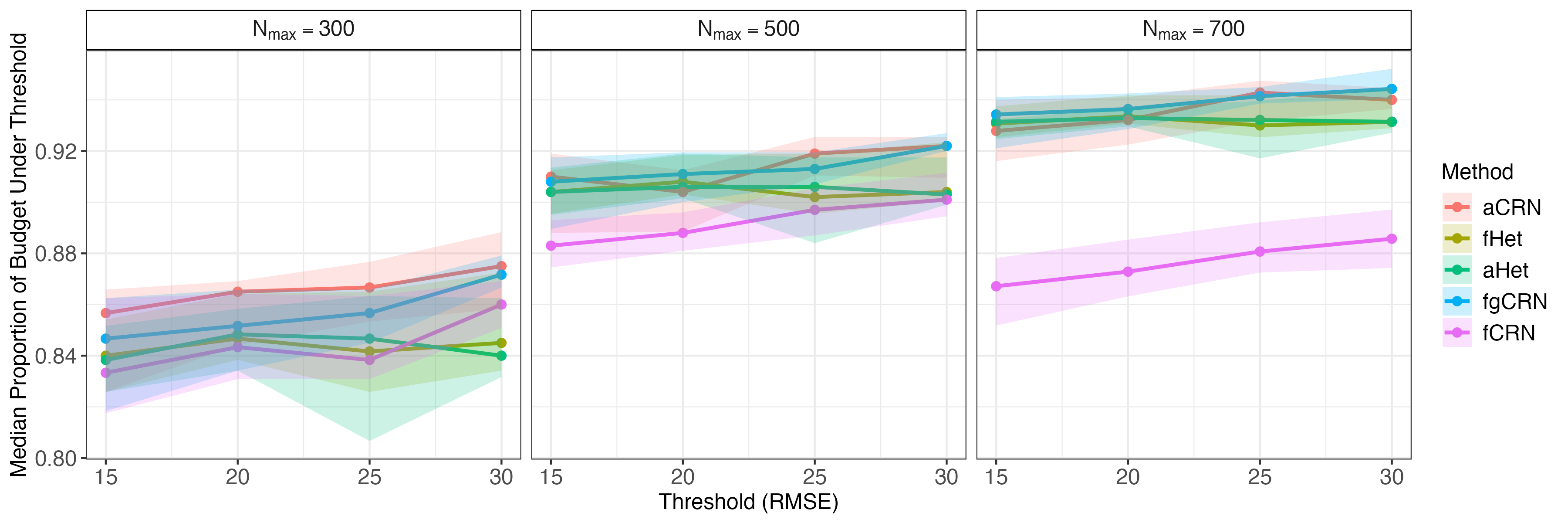}
    \caption{Means and standard errors of proportions of below-threshold trajectories found for the best-performing experimental design for each method.}
    \label{fig:pct_found_best_0.3_0.8}
\end{figure}

\begin{figure}[!ht]
    \centering
    \includegraphics[width=\textwidth]{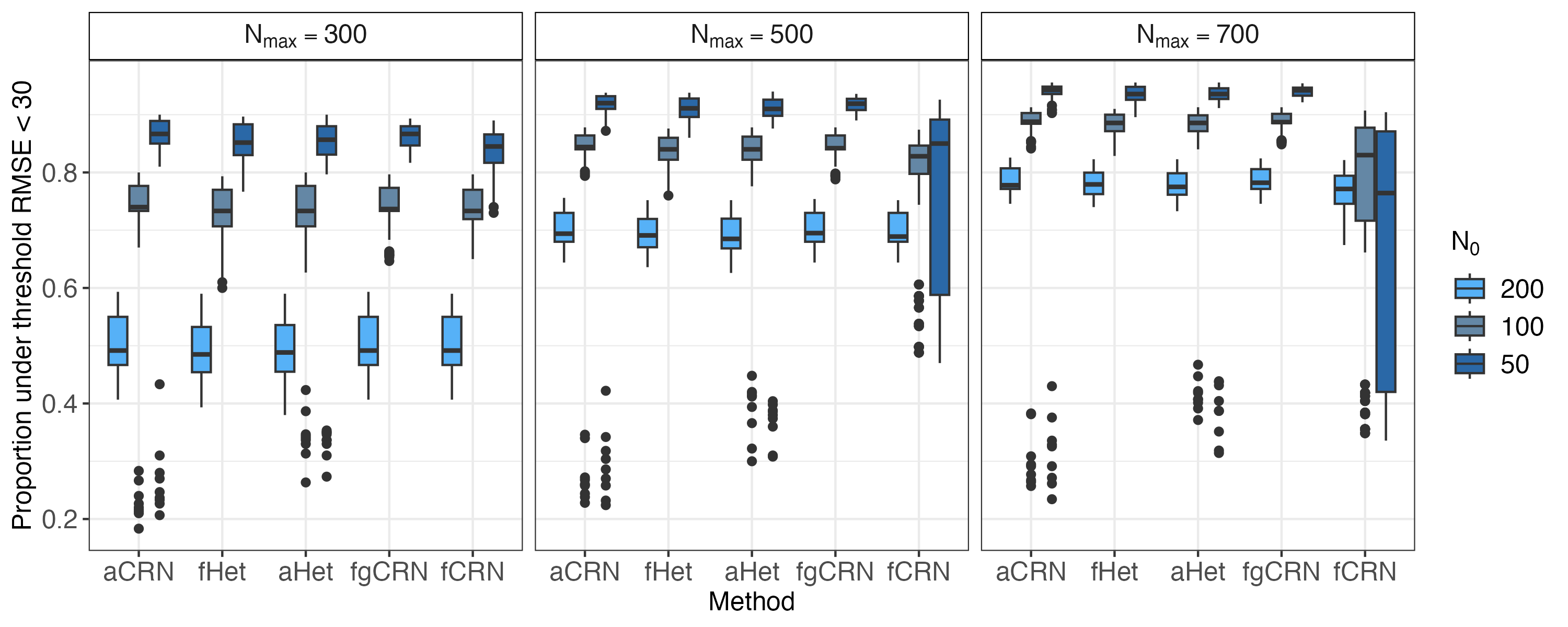}
    \caption{Proportion of trajectories with RMSE below 30 across all experiments varying the hyper-parameters, stratified by simulation budget $N_{\max}$, and initial design size $N_0$.}
    \label{fig:initial-design_0.3_0.8}
\end{figure}

\begin{figure}[!ht]
    \centering
    \includegraphics[width=\textwidth]{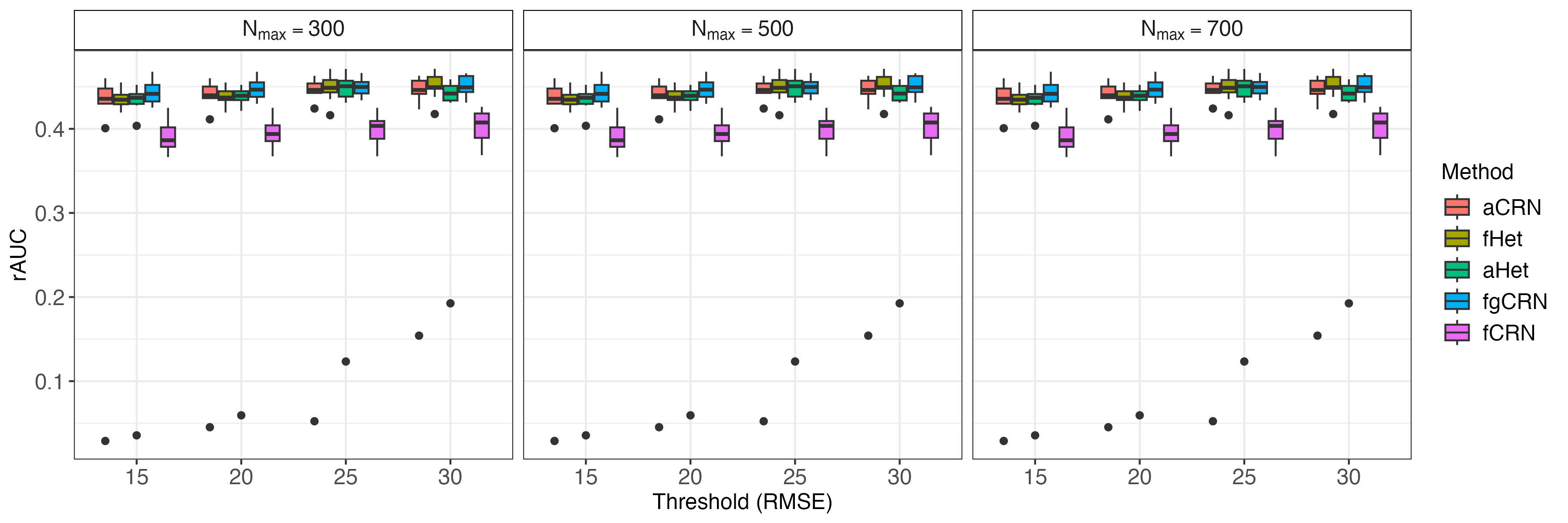}
    \caption{rAUC of best performing design for each method under different budget and varying thresholds.}
    \label{fig:auc_comparison_0.3_0.8}
\end{figure}


\end{document}